# ON THE DECIDABILITY AND COMPLEXITY OF METRIC TEMPORAL LOGIC OVER FINITE WORDS

JOËL OUAKNINE AND JAMES WORRELL

Oxford University Computing Laboratory, Oxford, UK
*e-mail address*: {joel,jbw}@comlab.ox.ac.uk

ABSTRACT. *Metric Temporal Logic (MTL)* is a prominent specification formalism for real-time systems. In this paper, we show that the satisfiability problem for MTL over finite timed words is decidable, with non-primitive recursive complexity. We also consider the model-checking problem for MTL: whether all words accepted by a given Alur-Dill timed automaton satisfy a given MTL formula. We show that this problem is decidable over finite words. Over infinite words, we show that model checking the safety fragment of MTL—which includes invariance and time-bounded response properties—is also decidable. These results are quite surprising in that they contradict various claims to the contrary that have appeared in the literature.

1. INTRODUCTION

In the linear-temporal-logic approach to verification, an execution of a system is modelled by a sequence of states or events. This representation abstracts away from the precise times of observations, retaining only their relative order. Such an approach is inadequate to express specifications of systems whose correct behaviour depends on quantitative timing requirements. To address this deficiency, much work has gone into adapting linear temporal logic to the real-time setting; see, e.g., [6, 7, 9, 10, 24, 27, 32, 35].

Real-time logics feature explicit time references, typically by recording timestamps throughout computations. In this paper, we concentrate exclusively on the *dense-time*, or *real-time*, semantics, in which the timestamps are drawn from the set of real numbers.[1] An important distinction among real-time models is whether one adopts a state-based semantics [7, 21, 32] or an event-based semantics [16, 9, 10, 18, 19, 35]. In the former, an execution of a system is modelled by a function that maps each point in time to the state propositions that are true at that moment. In the latter, one records only a countable sequence of events, corresponding to changes in the discrete state of the system. The distinction between these two semantic models is discussed, among others, in [8, 18]. As we will explain, the main results of this paper crucially depend on our adoption of the event-based model.

---

[1]By contrast, in *discrete-time* settings timestamps are usually integers, which yields more tractable theories that however correspond less closely to physical reality [19, 5].

One of the earliest and most popular proposals for extending temporal logic to the real-time setting is to replace the temporal operators by time-constrained versions—see [8] and the references therein. *Metric Temporal Logic (MTL)*, introduced fifteen years ago by Koymans [24], is a prominent and successful instance of this approach.[2] MTL extends Linear Temporal Logic by constraining the temporal operators by (bounded or unbounded) intervals of the real numbers. For example, the formula $\Diamond_{[3,4]} \varphi$ means that $\varphi$ will hold within 3 to 4 time units from now.

Unfortunately, over the state-based semantics, the satisfiability and model-checking problems for MTL are undecidable [16]. This has led some researchers to consider various restrictions on MTL to recover decidability; see, e.g., [6, 7, 19, 35]. Undecidability arises from the fact that MTL formulas can capture the computations of a Turing machine: configurations of the machine can be encoded within a single unit-duration time interval, since the density of time can accommodate arbitrarily large amounts of information. An MTL formula can then specify that configurations be accurately propagated from one time interval to the next, in such a way that the timed words satisfying the formula correspond precisely to the halting computations of the Turing machine.

It turns out that the key ingredient required for this procedure to go through is *punctuality*: the ability to specify that a particular event is always followed exactly one time unit later by another one: $\Box(p \rightarrow \Diamond_{=1} q)$. It has in fact been claimed that, in the state-based and the event-based semantics alike, any logic strong enough to express the above requirement will automatically be undecidable—see [8, 9, 18, 20], among others. While the claim is correct over the state-based semantics, we show in this paper that it is erroneous in the event-based semantics. Indeed, we show that both satisfiability and model checking for MTL over finite timed words are decidable, albeit with non-primitive recursive complexity. Over infinite words, we show that model checking the safety fragment of MTL—which includes invariance and punctual time-bounded response properties—is also decidable.

Upon careful analysis, one sees that the undecidability argument breaks down because, over the event-based semantics, MTL is only able to encode faulty Turing machines, that is, Turing machines suffering from *insertion errors*: while the formula $\Box(p \leftrightarrow \Diamond_{=1} q)$ ensures that every $p$ is followed exactly one time unit later by a $q$, there might be some $q$'s that were *not* preceded one time unit earlier by a $p$ (indeed, by any event at all). Intuitively, this problem does not occur over the state-based semantics because there the system is assumed to be under observation at all instants in time, and therefore any insertion error will automatically be detected thanks to the above formula.

MTL is also genuinely undecidable over the event-based semantics if in addition *past* temporal operators are allowed [9, 16]. Indeed, in this setting insertion errors can be detected by going backwards in time, and MTL formulas are therefore able to precisely capture the computations of perfect Turing machines.[3]

The decidability results that we present in this paper are obtained by translating MTL formulas into *timed alternating automata*. These generalise Alur-Dill timed automata, and, unlike the latter, are closed under complementation. Building on some of our previous work [28], using the theory of *well-structured transition systems*, we show that the language emptiness problem for one-clock timed alternating automata over finite timed words is decidable, which then entails the decidability of MTL satisfiability over finite timed words. We furthermore show how to extend these results to the model-checking problems discussed earlier. In addition, we show that MTL formulas can capture the computations of insertion channel machines; then, using a result of Schnoebelen about the complexity of reachability for lossy channel machines [33], we give a non-recursive primitive lower bound for the complexity of MTL satisfiability.

---

[2]As of early 2006, `http://scholar.google.com` lists over three hundred and fifty papers on the subject!

[3]The original undecidability proof in [9] was carried out in a monadic first-order theory of timed words, which subsumes both forward and past temporal operators.

1.1. **Related Work.** Existing decidability results for MTL involve placing restrictions on the semantics or the syntax of the logic to circumvent the problem of punctuality. Alur and Henzinger [9] showed that the satisfiability and model-checking problems for MTL relative to a discrete-time semantics are EXPSPACE-complete. Alur, Feder, and Henzinger [6, 7] introduced *Metric Interval Temporal Logic (MITL)* as a fragment of MTL in which the temporal operators may only be constrained by *nonsingular* intervals. They showed that the satisfiability and model-checking problems for MITL relative to a dense-time semantics are also EXPSPACE-complete. Wilke [35] considered MTL over a dense-time semantics with *bounded variability*, i.e., the semantics is parameterised by a bound $k$ on the number of events per unit time interval. He showed that the satisfiability problem is decidable in this semantics and that MTL with existential quantification over propositions is equally expressive as Alur-Dill timed automata.

A notion of timed alternating automaton very similar to the one considered here has recently and independently been introduced by Lasota and Walukiewicz [25]. They also prove that the finite-word language emptiness problem is decidable for one-clock timed alternating automata, and likewise establish a non-primitive recursive complexity bound for this procedure. However they do not consider any questions related to MTL, or timed logics in general.

A class of timed alternating tree automata has been defined by Dickhöfer and Wilke [14] in the context of model checking a real-time version of Computation Tree Logic, called TCTL. The language-emptiness problem for these automata is undecidable in general. However, TCTL model checking reduces to a special case of language emptiness, which is shown in [14] to be decidable using Alur and Dill's clock regions construction. In contrast, bounded-dimension clock regions do not suffice in the present paper: we combine clock regions with the notion of well-quasi-orders.

Another closely related paper is that of Abdulla and Jonsson [4] on networks of one-clock timed processes. This has a similar flavour to the work presented here in that it uses abstractions based on clock regions and also Higman's Lemma. The problems they study are however very different from the ones considered in this paper.

All the decidability results presented in this paper concern timed alternating automata over finite timed words, including the results that are ostensibly about infinite timed words. In particular, our model-checking procedure for the safety fragment of MTL over infinite timed words depends on the fact that any infinite timed word violating a safety property has a finite *bad prefix*, that is, a finite prefix none of whose extensions satisfies the property. Since writing the extended abstract of this paper [29], we have obtained some positive and negative decidability results about the language emptiness problem for timed alternating automata over infinite words. We discuss these results in the conclusion, Section 9.

## 2. Timed Words and Timed Automata

A *time sequence* $\tau = \tau_1 \tau_2 \tau_3 \ldots$ is a non-empty finite or infinite sequence of time values $\tau_i \in \mathbb{R}_{\geq 0}$ satisfying the following constraints (where $|\tau|$ denotes the length of $\tau$):

- *monotonicity*: $\tau_i \leq \tau_{i+1}$ for all $i$ such that $1 \leq i < |\tau|$
- *progress*: if $\tau$ is infinite, then $\{\tau_i : i \geq 1\}$ is unbounded.

A *timed word* over finite alphabet $\Sigma$ is a pair $\rho = (\sigma, \tau)$, where $\sigma = \sigma_1 \sigma_2 \sigma_3 \ldots$ is a non-empty finite or infinite word over $\Sigma$ and $\tau$ is a time sequence of the same length as $\sigma$. We also represent a timed word as a sequence of *timed events* by writing $\rho = (\sigma_1, \tau_1)(\sigma_2, \tau_2)(\sigma_3, \tau_3) \ldots$. Given a timed word $\rho = (\sigma, \tau)$ and $n \leq |\rho|$, let $\rho[1 \ldots n]$ denote the prefix $(\sigma_1, \tau_1) \ldots (\sigma_n, \tau_n)$. Finally, write $T\Sigma^*$ for the set of finite timed words over alphabet $\Sigma$, and $T\Sigma^\omega$ for the set of infinite timed words over $\Sigma$.

The requirement that infinite timed words be progressive is sometimes called *non-Zenoness* or *finite variability*. It is equivalent to the requirement that an infinite number of events not occur in a finite amount of time. Note however that, unlike [35], we place no *a priori* bound on the number of events that can occur in a time interval of unit duration.

### 2.1. Timed Automata.
Definition 2.1 recalls the standard notion of a *timed automaton* [5]. Elsewhere in this paper we refer to the timed automata defined below as *Alur-Dill automata*. This is to distinguish them from the more general class of *timed alternating automata*, which we introduce in Section 3 and which is our primary focus.

Let $X = \{x_1, \ldots, x_n\}$ be a finite set of *clock variables*. Define the set $\Phi_X$ of clock constraints over $X$ by the grammar

$$\varphi ::= \textbf{true} \mid x \bowtie c \mid \varphi_1 \wedge \varphi_2,$$

where $c \in \mathbb{N}$ is a non-negative integer, $x \in X$, and $\bowtie \in \{<, \leq, \geq, >\}$.

**Definition 2.1.** A *timed automaton* is a tuple $\mathcal{A} = (\Sigma, S, s_0, F, X, \Delta)$, where
- $\Sigma$ is a finite alphabet of events
- $S$ is a finite set of locations
- $s_0 \in S$ is an initial location
- $F \subseteq S$ is a set of accepting locations
- $X$ is a finite set of clock variables
- $\Delta \subseteq S \times \Sigma \times S \times \Phi_X \times 2^X$ is a finite set of edges. An edge $(s, a, s', \varphi, R)$ denotes an $a$-labelled transition from $s$ to $s'$, with precondition $\varphi$, and with the postcondition that all clocks in $R$ are reset to zero while all other clocks remain unchanged.

Given a timed automaton $\mathcal{A}$, let $c_{\max}$ be the maximum constant appearing in a clock constraint in $\mathcal{A}$. The set of clock values appropriate to $\mathcal{A}$ is defined to be $\mathsf{Val} = [0, c_{\max}] \cup \{\top\}$, where $\top$ represents any clock value strictly greater than $c_{\max}$. $\top$ satisfies the expected arithmetic and order-theoretic properties: if $v \in [0, c_{\max}]$ and $t \in \mathbb{R}_+$ are such that $v + t > c_{\max}$ then we write $v + t = \top$; we also define $\top + t = \top$ for all $t \in \mathbb{R}_{\geq 0}$; finally, we define $\top > v$ for all $v \in \mathsf{Val}$.[4] A *clock valuation* of $\mathcal{A}$ is a vector $\mathbf{v} = (v_1, \ldots, v_n)$, where $v_i \in \mathsf{Val}$ gives the value of clock $x_i$. If $t \in \mathbb{R}_{\geq 0}$, we let $\mathbf{v} + t$ be the clock valuation whose $i$-th component is $v_i + t$. A *state* of $\mathcal{A}$ is a pair $(s, \mathbf{v})$, where $s \in S$ is a location and $\mathbf{v}$ is a clock valuation. Write $Q = S \times \mathsf{Val}^n$ for the set of states of $\mathcal{A}$.

---

[4] Identifying all clock values strictly greater than $c_{\max}$ is harmless since such values are indistinguishable by clock constraints. Moreover this identification will later turn out to be technically advantageous.

Automaton $\mathcal{A}$ induces a labelled transition system $\mathcal{T}_\mathcal{A} = (Q, \rightsquigarrow, \longrightarrow)$ on the set of states, where $\rightsquigarrow \subseteq Q \times \mathbb{R}_{\geq 0} \times Q$ is called the *delay-step relation*, and $\longrightarrow \subseteq Q \times \Sigma \times Q$ is called the *discrete-step relation*. Delay steps model the evolution of time while the automaton remains in a given location, while discrete steps correspond to instantaneous transitions between locations. The delay-step transition relation is deterministic, and is defined by $(s, \mathbf{v}) \stackrel{t}{\rightsquigarrow} (s, \mathbf{v} + t)$, where $t \in \mathbb{R}_{\geq 0}$. The discrete-step relation is defined by $(s, \mathbf{v}) \stackrel{a}{\longrightarrow} (s', \mathbf{v}')$ iff there is an edge $(s, a, s', \varphi, R) \in \Delta$ such that $\mathbf{v}$ satisfies $\varphi$, $v'_i = 0$ for all $x_i \in R$ and $v'_i = v_i$ for all $x_i \notin R$.

Let $\rho = (\sigma, \tau)$ be a timed word, and write $d_i = \tau_i - \tau_{i-1}$ for the time delay between the $(i-1)$th and $i$th events of $\rho$, where $1 \leq i \leq |\rho|$, and, by convention, $\tau_0 = 0$. Define a *run* of $\mathcal{A}$ on $\rho$ to be an alternating sequence of time delays and discrete steps in $\mathcal{T}_\mathcal{A}$:

$$(s_0, \mathbf{v}_0) \stackrel{d_1}{\rightsquigarrow} (s_1, \mathbf{v}_1) \stackrel{\sigma_1}{\longrightarrow} (s_2, \mathbf{v}_2) \stackrel{d_2}{\rightsquigarrow} (s_3, \mathbf{v}_3) \stackrel{\sigma_2}{\longrightarrow} \cdots \stackrel{d_n}{\rightsquigarrow} (s_{2n-1}, \mathbf{v}_{2n-1}) \stackrel{\sigma_n}{\longrightarrow} \cdots,$$

where $s_0$ is the initial location and $\mathbf{v}_0$ maps every clock variable to 0.

A finite run is *accepting* if the last location in the run is accepting. An infinite run is accepting if infinitely many control states in the run are accepting. We write $L_f(\mathcal{A})$ for the set of finite timed words over which $\mathcal{A}$ has an accepting run, and we write $L_\omega(\mathcal{A})$ for the set of infinite timed words over which $\mathcal{A}$ has an accepting run.

## 3. Timed Alternating Automata

In this section we define timed alternating automata. These arise by extending alternating automata [11, 13, 34] with clock variables, in much the same way that Alur-Dill timed automata extend nondeterministic finite automata. A similar notion has independently been investigated by Lasota and Walukiewicz in a recent paper [25]. It will soon become apparent that timed alternating automata strictly generalise Alur-Dill automata. However we chose to introduce Alur-Dill automata separately, in Section 2, since by so doing we can avoid considering timed alternating automata with Büchi acceptance conditions. (This greatly simplifies the definition of a run of an alternating automaton because we can elide the tree structure—see below.)

Timed alternating automata can in general be defined to have any number of clocks. Our goal, however, is to use them to represent metric temporal logic formulas, for which one clock suffices. Accordingly, we shall exclusively focus on one-clock timed alternating automata in this paper.[5] In this section we only consider timed alternating automata over *finite* timed words.

### 3.1. One-clock Timed Alternating Automata. 
Let $S$ be a finite set of *locations* and let $x$ be a distinguished clock variable. We define a set of formulas $\Phi(S)$ by the grammar:

$$\varphi ::= \mathbf{true} \mid \mathbf{false} \mid \varphi_1 \wedge \varphi_2 \mid \varphi_1 \vee \varphi_2 \mid s \mid x \bowtie c \mid x.\varphi,$$

where $c \in \mathbb{N}$, $\bowtie \in \{<, \leq, \geq, >\}$, and $s \in S$. A term of the form $x \bowtie c$ is called a *clock constraint*, whereas the expression $x.\varphi$ is a binding construct corresponding to the operation of resetting the clock $x$ to 0.

---

[5]We note in passing that virtually all decision problems, and in particular language emptiness, are undecidable for timed alternating automata that have more than one clock; cf. Section 4.

In the definition of a timed alternating automaton, below, the transition function maps each location $s \in S$ and event $a \in \Sigma$ to an expression in $\Phi(S)$. Thus alternating automata allow two modes of branching: existential branching, represented by disjunction, and universal branching, represented by conjunction.

**Definition 3.1.** A timed alternating automaton is a tuple $\mathcal{A} = (\Sigma, S, s_0, F, \delta)$, where
- $\Sigma$ is a finite alphabet
- $S$ is a finite set of locations
- $s_0 \in S$ is the initial location
- $F \subseteq S$ is a set of accepting locations
- $\delta : S \times \Sigma \to \Phi(S)$ is the transition function.

The notion of a run of a timed alternating automaton, defined below, is somewhat involved, so we first give an example.

**Example 3.2.** We define an automaton $\mathcal{A}$ over the singleton alphabet $\Sigma = \{a\}$ that accepts all those finite timed words in which no two events are separated by exactly one time unit. This language is known not to be expressible as the language of an Alur-Dill timed automaton [22]. The required timed alternating automaton has set of locations $\{s_0, s_1\}$, with $s_0$ initial, and both $s_0$ and $s_1$ accepting. The transition function is defined by:

$$\delta(s_0, a) = s_0 \wedge x.s_1$$
$$\delta(s_1, a) = s_1 \wedge x \neq 1 \, .$$

A run of $\mathcal{A}$ starts in location $s_0$. Every time an $a$-event occurs, the automaton makes a conjunctive transition to both $s_0$ and $s_1$, thus opening up a new thread of computation. The automaton resets a fresh copy of clock $x$ whenever it transitions from location $s_0$ to $s_1$, and the transition rule for $s_1$ ensures that no event can happen when the value of this clock equals one. Every run of this automaton is accepting, since every location is accepting, but there is no run over any word in which two events are separated by exactly one time unit.

We now proceed to the formal definitions. Let $c_{\max}$ be the maximum constant mentioned in the definition of the transition function of $\mathcal{A}$, and, as with Alur-Dill automata, define the set of clock values relevant to $\mathcal{A}$ to be $\mathsf{Val} = [0, c_{\max}] \cup \{\top\}$. A *state* of $\mathcal{A}$ is a pair $(s, v)$, where $s \in S$ is a location and $v \in \mathsf{Val}$ is a *clock valuation*. Write $Q = S \times \mathsf{Val}$ for the set of all states of $\mathcal{A}$.

A set of states $M \subseteq Q$ and a clock valuation $v \in \mathsf{Val}$ defines a Boolean valuation on $\Phi(S)$ as follows:
- $M \models_v \mathbf{true}$
- $M \models_v \varphi_1 \wedge \varphi_2$ if $M \models_v \varphi_1$ and $M \models_v \varphi_2$
- $M \models_v \varphi_1 \vee \varphi_2$ if $M \models_v \varphi_1$ or $M \models_v \varphi_2$
- $M \models_v s$ if $(s, v) \in M$
- $M \models_v x \bowtie c$ if $v \bowtie c$
- $M \models_v x.\varphi$ if $M \models_0 \varphi$.

We say that a set of states $M$ is a *minimal model* of a formula $\varphi \in \Phi(S)$ with respect to clock value $v \in \mathsf{Val}$ if $M \models_v \varphi$ and there is no proper subset $M' \subset M$ with $M' \models_v \varphi$.[6]

---
[6] Our use of minimal models here is a technical convenience, since, as we will see later, the minimal models of formula $\varphi$ can be directly related to the syntactic structure of $\varphi$ when the latter is given in disjunctive normal form.

**Example 3.3.** The minimal models of $\varphi \equiv x.s_0 \wedge (s_1 \vee s_2)$ with respect to the clock value $v = 1.2$ are $\{(s_0, 0), (s_1, 1.2)\}$ and $\{(s_0, 0), (s_2, 1.2)\}$.

A *configuration* of $\mathcal{A}$ is a finite set of states; thus the set of configurations is the finite powerset of the set $Q$ of states, and is denoted $\wp(Q)$. The *initial configuration* is $\{(s_0, 0)\}$ and a configuration is *accepting* if every location that it contains is accepting. Note in particular that the empty configuration is accepting. Given a configuration $C$ and a time delay $t \geq 0$, denote by $C + t$ the configuration $\{(s, v+t) : (s, v) \in C\}$.

The language accepted by a timed alternating automaton over finite words can be described in terms of a transition system of configurations, defined below.

**Definition 3.4.** Given a timed alternating automaton $\mathcal{A}$, we define the labelled transition system $\mathcal{T}_\mathcal{A} = (\wp(Q), \rightsquigarrow, \longrightarrow)$ over the set of configurations as follows. The $(\mathbb{R}_{\geq 0})$-labelled transition relation $\rightsquigarrow \;\subseteq\; \wp(Q) \times \mathbb{R}_{\geq 0} \times \wp(Q)$ captures time evolutions, or *delay steps*, and is defined by $C \stackrel{t}{\rightsquigarrow} C'$ iff $C' = C + t$.

The $\Sigma$-labelled transition relation $\longrightarrow \;\subseteq\; \wp(Q) \times \Sigma \times \wp(Q)$ captures instantaneous transitions between locations, or *discrete steps*. Let $C = \{(s_i, v_i)\}_{i \in I}$. We include a transition $C \stackrel{a}{\longrightarrow} C'$ iff one can write $C' = \bigcup_{i \in I} M_i$, where, for each $i \in I$, $M_i$ is a minimal model of $\delta(s_i, a)$ with respect to $v_i$.

Let $\rho = (\sigma, \tau)$ be a finite timed word with $|\rho| = n$, and write $d_i = \tau_i - \tau_{i-1}$ for $1 \leq i \leq |\rho|$, where, by convention, $\tau_0 = 0$. Define a *run* of $\mathcal{A}$ on $\rho$ to be a finite alternating sequence of time delays and discrete steps in $\mathcal{T}_\mathcal{A}$:

$$C_0 \stackrel{d_1}{\rightsquigarrow} C_1 \stackrel{\sigma_1}{\longrightarrow} C_2 \stackrel{d_2}{\rightsquigarrow} C_3 \stackrel{\sigma_2}{\longrightarrow} \cdots \stackrel{d_n}{\rightsquigarrow} C_{2n-1} \stackrel{\sigma_n}{\longrightarrow} C_{2n} ,$$

where $C_0$ is the initial configuration. We say that the run is *accepting* if the last configuration $C_{2n}$ is accepting, and we say that the timed word $\rho$ is *accepted* by $\mathcal{A}$ if there is some accepting run of $\mathcal{A}$ on $\rho$. We write $L_f(\mathcal{A}) \subseteq T\Sigma^*$ for the language of finite timed words accepted by $\mathcal{A}$.[7]

**Example 3.5.** A time-bounded response property such as 'for every $a$-event there is a $b$-event exactly one time unit later' can be expressed by the following automaton. Let $\mathcal{A}$ have two locations $\{s_0, s_1\}$ with $s_0$ the initial and only accepting location, and transition function $\delta$ given by the following table:

|       | $a$              | $b$              |
|-------|------------------|------------------|
| $s_0$ | $s_0 \wedge x.s_1$ | $s_0$            |
| $s_1$ | $s_1$            | $(x = 1) \vee s_1$ |

Location $s_0$ represents an invariant, and is present in every configuration in any run of $\mathcal{A}$. When an $a$-event occurs, the conjunction in the definition of $\delta(s_0, a)$ results in the creation of a new thread of computation, starting in location $s_1$. Since this location is not accepting, the automaton must eventually leave it. This is only possible if a $b$-event happens exactly one time unit after the new thread was spawned.

---

[7]It is usual to define a run of an alternating automaton to be a *tree* of states. However, over finite words the branching structure plays no role in the definition of acceptance, and we simply define a run to be a sequence of configurations, where each configuration represents a given level of the run tree.

3.2. **Duality and Complementation.** The following derivation shows that the class of languages definable by timed alternating automata is closed under complement. Since it is straightforward to show that this class is also closed under union, timed alternating automata are closed under all Boolean operations. The arguments presented here are similar to the untimed case [11, 13].

Given $\varphi \in \Phi(S)$, we define the *dual formula* $\overline{\varphi} \in \Phi(S)$ as follows. The dual of a clock constraint is its negation (e.g., $\overline{x < k} = x \geq k$), whereas each location is self-dual: $\overline{s} = s$ for $s \in S$. For the propositional connectives we have the usual de Morgan dualities: $\overline{\mathbf{true}} = \mathbf{false}$, $\overline{\mathbf{false}} = \mathbf{true}$, $\overline{\varphi_1 \vee \varphi_2} = \overline{\varphi_1} \wedge \overline{\varphi_2}$ and $\overline{\varphi_1 \wedge \varphi_2} = \overline{\varphi_1} \vee \overline{\varphi_2}$. Finally, clock resets distribute through the duality operator: $\overline{x.\varphi} = x.\overline{\varphi}$.

Let $\mathcal{A} = (\Sigma, S, s_0, F, \delta)$ be an alternating timed automaton, and denote by $Q$ its set of states. The complement automaton $\mathcal{A}^c$ is defined by $\mathcal{A}^c = (\Sigma, S, s_0, S \setminus F, \overline{\delta})$, where $\overline{\delta}(s, a) = \overline{\delta(s, a)}$ for each $s \in S$ and $a \in \Sigma$. Thus we take the dual transition function and the complement of the set of accepting locations.

**Proposition 3.6.** *Let $\varphi \in \Phi(S)$ be a formula over set of locations $S$ and let $v \in \mathsf{Val}$ be a clock valuation. Given a set of states $P \subseteq Q$ we have $P \models_v \varphi$ iff $Q \setminus P \not\models_v \overline{\varphi}$.*

*Proof.* The proof is by structural induction on $\varphi$, and is straightforward from the definition of $\overline{\varphi}$. □

**Proposition 3.7.** $L_f(\mathcal{A}) \cap L_f(\mathcal{A}^c) = \emptyset$.

*Proof.* Suppose that both $\mathcal{A}$ and $\mathcal{A}^c$ have runs on the same timed word $\rho = (\sigma, \tau)$, with $|\rho| = n$. Denote the run of $\mathcal{A}$ by

$$C_0 \stackrel{d_1}{\leadsto} C_1 \stackrel{\sigma_1}{\longrightarrow} C_2 \stackrel{d_2}{\leadsto} C_3 \stackrel{\sigma_2}{\longrightarrow} \cdots \stackrel{d_n}{\leadsto} C_{2n-1} \stackrel{\sigma_n}{\longrightarrow} C_{2n} ,$$

and denote the run of $\mathcal{A}^c$ by

$$D_0 \stackrel{d_1}{\leadsto} D_1 \stackrel{\sigma_1}{\longrightarrow} D_2 \stackrel{d_2}{\leadsto} D_3 \stackrel{\sigma_2}{\longrightarrow} \cdots \stackrel{d_n}{\leadsto} D_{2n-1} \stackrel{\sigma_n}{\longrightarrow} D_{2n} .$$

We show by induction on $i \leq 2n$ that $C_i \cap D_i$ is non-empty. In particular, we deduce that $C_{2n}$ and $D_{2n}$ meet, so the two runs cannot both be accepting since $\mathcal{A}$ and $\mathcal{A}^c$ have disjoint sets of accepting states.

The base case of the induction is just the observation that $C_0 = D_0 = \{(s_0, 0)\}$. For the induction step, suppose that $(s, v) \in C_i \cap D_i$. In case $i = 2j$ is even, that is, the next transition is a time delay, then $(s, v + d_{j+1}) \in C_{i+1} \cap D_{i+1}$. In case $i = 2j + 1$ is odd, that is, the next transition is a discrete step, then $C_{i+1} \models_v \delta(s, \sigma_{j+1})$ and $D_{i+1} \models_v \overline{\delta(s, \sigma_{j+1})}$. It follows from Proposition 3.6 that $C_{i+1}$ and $D_{i+1}$ are not disjoint. This completes the induction step. □

**Proposition 3.8.** $L_f(\mathcal{A}) \cup L_f(\mathcal{A}^c) = T\Sigma^*$.

*Proof.* We claim that, given a finite timed word $\rho = (\sigma, \tau)$ and a set of states $P \subseteq Q$, either $\mathcal{A}$ has a run on $\rho$ whose last configuration is a subset of $P$, or $\mathcal{A}^c$ has a run on $\rho$ whose last configuration is a subset of $Q \setminus P$. The proposition follows from the claim by taking $P$ to be the set of states in $\mathcal{A}$ whose underlying location is accepting.

We prove the claim by induction on $|\rho|$ as follows. Let $\rho = (\sigma, \tau)$ and $P \subseteq Q$ be given as in the claim, with $|\rho| = n + 1$. Also, let $d_{n+1} = \tau_{n+1} - \tau_n$ and write

$$pred(P) = \{(s, v) \in Q : P \models_{v + d_{n+1}} \delta(s, \sigma_{n+1})\} .$$

Observe also that by Proposition 3.6
$$Q \setminus pred(P) = \{(s,v) \in Q : P \not\models_{v+d_{n+1}} \delta(s, \sigma_{n+1})\}$$
$$= \{(s,v) \in Q : Q \setminus P \models_{v+d_{n+1}} \overline{\delta(s, \sigma_{n+1})}\}. \quad (3.1)$$

By induction, either $\mathcal{A}$ has a run on $\rho[1\ldots n]$ whose last configuration $C$ is a subset of $pred(P)$, or $\mathcal{A}^c$ has a run on $\rho[1\ldots n]$ whose last configuration $D$ is a subset of $Q \setminus pred(P)$. In the former case, it is immediate that we can extend the given run of $\mathcal{A}$ into a run on $\rho$. Indeed, since $C \subseteq pred(P)$, for each $(s,v) \in C$ we can choose a finite subset of $P$ that is a minimal model of $\delta(s, \sigma_{n+1})$ with respect to clock value $v + d_{n+1}$. In the latter case, in similar fashion, it follows from (3.1) that $\mathcal{A}^c$ has a run on $\rho$ whose last configuration is a subset of $Q \setminus P$. □

**Corollary 3.9.** *The class of languages definable by timed alternating automata is effectively closed under all Boolean operations.*

## 4. Decidability of Language Emptiness

It is well known that the universality problem for Alur-Dill timed automata is undecidable [5]. In fact the proof in [5] shows undecidability for the subclass of Alur-Dill automata with at most two clocks. Since the class of timed alternating automata is closed under complement and includes the class of Alur-Dill automata, it immediately follows that the language-emptiness problem for two-clock timed alternating automata is undecidable. However we show in this section that if we restrict to alternating automata with a single clock, then language emptiness is decidable. The decision procedure that we present is a generalisation of the algorithm for deciding universality for one-clock Alur-Dill automata that appeared in the extended abstract [28].

In the remainder of this section we assume that $\mathcal{A} = (\Sigma, S, s_0, \delta, F)$ is a one-clock alternating automaton, and we denote by $Q$ the set of states of this automaton. The language-emptiness problem for $\mathcal{A}$ is equivalent to the following reachability question on the derived transition system $\mathcal{T}_\mathcal{A}$: 'Is there a path from the initial configuration to an accepting configuration?'. However it is not immediate how to decide this question since $\mathcal{T}_\mathcal{A}$ has uncountably many states: indeed each state has uncountably many successors under the delay-step relation.

4.1. **The Bisimulation Lemma.** In this section we isolate a sub-transition-system of $\mathcal{T}_\mathcal{A}$, denoted $\mathcal{W}_\mathcal{A}$, that is effective and is, in a certain sense, bisimilar to $\mathcal{T}_\mathcal{A}$.[8] In particular, $\mathcal{W}_\mathcal{A}$ has only countably many states and is finitely branching. Moreover the state space of $\mathcal{W}_\mathcal{A}$ possesses an effective well-quasi-order, which we use to prove termination of our reachability algorithm.

Recall that the set of clock values relevant to $\mathcal{A}$ is $\mathsf{Val} = [0, c_{\max}] \cup \{\top\}$. Define the *fractional part* of $v \in \mathsf{Val} \setminus \{\top\}$ by $frac(v) = v - \lfloor v \rfloor$ (where $\lfloor \cdot \rfloor$ denotes the floor function). It is also technically convenient to define $frac(\top) = 0$.

---

[8]In the extended abstract of this paper $\mathcal{W}_\mathcal{A}$ was described as a *quotient* of $\mathcal{T}_\mathcal{A}$, akin to the clock-region quotient of an Alur-Dill automaton. However in our opinion the technical details are more straightforward under the present approach.

**Definition 4.1** (Clock Regions). Define an equivalence relation $\sim$ on the set of clock values Val by $u \sim v$ if either $u, v = \top$, or $u, v \neq \top$, $\lceil u \rceil = \lceil v \rceil$ and $\lfloor u \rfloor = \lfloor v \rfloor$ (where $\lceil \cdot \rceil$ denotes ceiling). The corresponding set of equivalence classes, or *regions*, is $REG = \{r_0, r_1, \ldots, r_{2c_{\max}+1}\}$, where $r_{2i} = \{i\}$ for $i \leq c_{\max}$, $r_{2i+1} = (i, i+1)$ for $i < c_{\max}$, and $r_{2c_{\max}+1} = \{\top\}$. Let $reg(v)$ denote the equivalence class of $v \in \mathsf{Val}$.

The intuition behind the transition system $\mathcal{W}_\mathcal{A}$ is that we can ignore those time delays in $\mathcal{T}_\mathcal{A}$ that leave unchanged the regions of the clock values in a configuration. We only consider time delays that take a configuration to its *time successor*:

**Definition 4.2.** Let $C \subseteq Q$ be an $\mathcal{A}$-configuration. If $C$ is non-empty then define $\mu = \max\{frac(v) : (s, v) \in C\}$ to be the maximum fractional part of the clock values appearing in $C$. Now define the *time successor* of $C$ to be the configuration $next(C)$ given by the following clauses:
- if $C = \emptyset$ then $next(C) = C$
- if $(s, v) \in C$ for some integer clock value $v \in [0, c_{\max}]$ then $next(C) = C + (1 - \mu)/2$
- if neither of the above cases hold then $next(C) = C + (1 - \mu)$.

**Example 4.3.** Suppose that the maximum constant appearing in $\mathcal{A}$ is $c_{\max} = 3$. Consider a configuration $C = \{(s, 1.25), (t, 2.5), (s, 0.75)\}$. Then $next(C) = \{(s, 1.5), (t, 2.75), (s, 1)\}$ (in which time has advanced by 0.25 units, and the clock value in $C$ with largest fractional part has moved to a new region while all other clock values remain in the same regions). On the other hand, if $C = \{(s, 1), (t, 0.5)\}$, then $next(C) = \{(s, 1.25), (t, 0.75)\}$ (in which the clock value in $C$ with fractional part zero moves to a new region, while all other clock values remain in the same regions). Finally, the time successor of $C = \{(s, 0.5), (t, 3)\}$ is $\{(s, 0.75), (t, \top)\}$.

**Definition 4.4.** Define the labelled transition system $\mathcal{W}_\mathcal{A}$ as follows.
- **Alphabet.** The alphabet of $\mathcal{W}_\mathcal{A}$ is $\Sigma \cup \{\varepsilon\}$.
- **States.** The states of $\mathcal{W}_\mathcal{A}$ are those $\mathcal{A}$-configurations $C \subseteq Q$ in which all clock values are rational (henceforth call such configurations rational).
- **Transitions.** Each state $C$ has a unique $\varepsilon$-transition to its time successor $next(C)$. For $a \in \Sigma$, we postulate that $C \xrightarrow{a} C'$ in $\mathcal{W}_\mathcal{A}$ if $C \xrightarrow{a} C'$ in $\mathcal{T}_\mathcal{A}$.

Thus $\mathcal{W}_\mathcal{A}$ differs from $\mathcal{T}_\mathcal{A}$ in containing only rational configurations and retaining only those delay steps between a configuration and its time successor (renaming these as $\varepsilon$-transitions). Next we show that $\mathcal{W}_\mathcal{A}$ and $\mathcal{T}_\mathcal{A}$ are bisimilar in a certain sense. To this end, we first reexamine the notion of the minimal model of a formula $\varphi \in \Phi(S)$ over the set of locations $S$ of $\mathcal{A}$ (cf. Section 3).

Any formula $\varphi \in \Phi(S)$ can be written in disjunctive normal form $\varphi \equiv \bigvee_{j \in J} \bigwedge A_j$, where each $A_j$ is a set of terms of the form $s$, $x.s$, and $x \bowtie c$ (which we call *atoms*). The minimal models of $\varphi$ can be read off from the disjunctive normal form as follows. For a set of atoms $A$ and a clock valuation $v \in \mathsf{Val}$, let $A[v] \subseteq Q$ be the set of states given by $A[v] = \{(s, v) : s \in A\} \cup \{(s, 0) : x.s \in A\}$. Then each minimal model $M$ of $\varphi$ with respect to $v$ has the form $M = A_j[v]$, for some $j \in J$, where $v$ satisfies all the clock constraints in $A_j$.

**Example 4.5.** For the formula $\varphi \equiv x.s_0 \wedge (s_1 \vee s_2)$ from Example 3.3, the equivalent disjunctive normal form is $(x.s_0 \wedge s_1) \vee (x.s_0 \wedge s_2)$. Then the two minimal models of $\varphi$ with respect to the clock value $v \in \mathsf{Val}$ are $(x.s_0 \wedge s_1)[v] = \{(s_0, 0), (s_1, v)\}$ and $(x.s_0 \wedge s_2)[v] = \{(s_0, 0), (s_1, v)\}$.

**Definition 4.6.** Define the relation $\equiv \; \subseteq \wp(Q) \times \wp(Q)$ by $C \equiv C'$ iff there is a bijection $f \colon C \to C'$ such that: (i) $f(s, u) = (t, u')$ implies $s = t$ and $u \sim u'$; (ii) if $f(s, u) = (s, u')$ and $f(t, v) = (t, v')$, then $frac(u) \leq frac(v)$ implies $frac(u') \leq frac(v')$.

**Lemma 4.7** (Bisimulation Lemma). *Suppose that $C$ and $D$ are $\mathcal{A}$-configurations such that $C \equiv D$. Then for each transition $C \xrightarrow{\alpha} C'$, with $\alpha \in \Sigma \cup \{\varepsilon\}$, there exists a configuration $D'$ and a transition $D \xrightarrow{\alpha} D'$ such that $C' \equiv D'$.*

*Proof.* Write $C = \{(s_i, u_i)\}_{i \in I}$ and $D = \{(t_i, v_i)\}_{i \in I}$, and let $f \colon C \to D$, given by $f(s_i, u_i) = (t_i, v_i)$, be the bijection witnessing $C \equiv D$.

*Matching $\Sigma$-labelled transitions.* Suppose $C$ makes a transition $C \xrightarrow{a} C'$ for some $a \in \Sigma$. By the above considerations on minimal models, we know that $C' = \bigcup_{i \in I} A_i[u_i]$, where, for each $i \in I$, the set of atoms $A_i$ is a clause in the disjunctive normal form expression for $\delta(s_i, a)$. Writing $D' = \bigcup_{i \in I} A_i[v_i]$, we have $D \xrightarrow{a} D'$. (Here we rely on the fact that $u_i \sim v_i$, so that $u_i$ and $v_i$ satisfy the same clock constraints.) Now the set of clock values appearing in $C'$ is a subset of $\{u_i : i \in I\} \cup \{0\}$ since $\Sigma$-labelled transitions either leave clocks unchanged or reset them to 0. Thus $C' \equiv D'$ since we can define a bijection $f' \colon C' \to D'$ by $f'(s, u_i) = (s, v_i)$ if $(s, u_i) \in C'$ and $f'(s, 0) = (s, 0)$ if $(s, 0) \in C'$.

*Matching $\varepsilon$-transitions.* Since each configuration makes a unique $\varepsilon$-transition to its time successor, we need only show that $next(C) \equiv next(D)$. Now $next(C)$ has the form $\{(s_i, u'_i)\}_{i \in I}$, where, for some time delay $t > 0$, $u'_i = u_i + t$; similarly $next(D)$ has the form $\{(t_i, v'_i)\}_{i \in I}$, where, for some time delay $t' > 0$, $v'_i = v_i + t'$. The effect of the time delay $t$ on $C$ is either to leave the order of the fractional parts of the clocks unchanged or to cyclically permute the order by one place so that the clock with greatest fractional part in $C$ has zero fractional part in $next(C)$. A similar statement holds for $D$. In any case, we have $u'_i \sim v'_i$ for each $i \in I$, and the bijection $f' : next(C) \to next(D)$ defined by $f'(s_i, u'_i) = (t_i, v'_i)$ witnesses $next(C) \equiv next(D)$. □

Let $\xrightarrow{\varepsilon^*}$ denote the reflexive transitive closure of the relation $\xrightarrow{\varepsilon}$. The following simple corollary of the Bisimulation Lemma shows that, up to $\equiv$-equivalence, there is no loss in expressiveness in replacing the delay-step transition relation with $\xrightarrow{\varepsilon^*}$.

**Corollary 4.8.** *Suppose that $C, D \subseteq Q$ are $\mathcal{A}$-configurations such that $C \equiv D$. Then for any time delay $C \overset{t}{\rightsquigarrow} C'$ there exists a configuration $D'$, with $D \xrightarrow{\varepsilon^*} D'$ and $C' \equiv D'$.*

*Proof.* Observe that $C \overset{t}{\rightsquigarrow} C'$ implies that $C' \equiv next^n(C)$ for some $n \geq 0$. From the Bisimulation Lemma we get that $next^n(C) \equiv next^n(D)$. The proposition follows by taking $D' = next^n(D)$. □

**Proposition 4.9.** *If configuration $C$ is reachable from the initial configuration $C_0$ in $\mathcal{T}_\mathcal{A}$, then there is a rational configuration $C'$, with $C \equiv C'$, such that $C'$ is reachable from $C_0$ in $\mathcal{W}_\mathcal{A}$.*

*Proof.* Given a path $C_0 \overset{d_1}{\rightsquigarrow} C_1 \xrightarrow{\sigma_1} C_2 \overset{d_2}{\rightsquigarrow} C_3 \xrightarrow{\sigma_2} \cdots \xrightarrow{\sigma_n} C_{2n}$ in $\mathcal{T}_\mathcal{A}$, we generate, step by step, a 'matching' path of rational configurations in $\mathcal{W}_\mathcal{A}$

$$C_0 \xrightarrow{\varepsilon^*} C'_1 \xrightarrow{\sigma_1} C'_2 \xrightarrow{\varepsilon^*} C'_3 \xrightarrow{\sigma_2} \cdots \xrightarrow{\sigma_n} C'_{2n}$$

such that $C_i \equiv C'_i$ for $0 < i \leq 2n$. Given $C'_i \equiv C_i$, if $i$ is odd we use the Bisimulation Lemma to generate the next configuration $C'_{i+1}$, and if $i$ is even we use Corollary 4.8 to generate $C'_{i+1}$. □

We have now reduced the language-emptiness problem for $\mathcal{A}$ to the following reachability question for $\mathcal{W}_\mathcal{A}$: 'Is there a path from the initial configuration to an accepting configuration?'. Although $\mathcal{W}_\mathcal{A}$ is simpler than $\mathcal{T}_\mathcal{A}$, it still has infinitely many states (indeed, even the quotient of $\mathcal{W}_\mathcal{A}$ by $\equiv$ is infinite-state, since $\equiv$ only relates configurations of the same cardinality). We circumvent this problem by exhibiting a *well-quasi-order* on the state space of $\mathcal{W}_\mathcal{A}$. This serves in lieu of finiteness to guarantee the termination of a state-exploration algorithm that computes a conservative over-approximation of the set of reachable states. This is described in the next subsection in terms of the theory of *well-structured transition systems* [15].

4.2. **Well-quasi-orders.** Recall that a quasi-order $(W, \sqsubseteq)$ consists of a set $W$ together with a reflexive, transitive relation $\sqsubseteq$. An infinite sequence $w_1, w_2, w_3, \ldots$ in $(W, \sqsubseteq)$ is said to be *saturating* if there exist indices $i < j$ such that $w_i \sqsubseteq w_j$. $(W, \sqsubseteq)$ is a *well-quasi-order (wqo)* if every infinite sequence in $(W, \sqsubseteq)$ is saturating.

We can extend a quasi-order $(W, \sqsubseteq)$ to a quasi-order $(W^*, \sqsubseteq)$ on the set of finite words over alphabet $W$ as follows. Define $w_1 \ldots w_m \sqsubseteq v_1 \ldots v_n$ if there exists a strictly increasing function $f : \{1 \ldots m\} \to \{1, \ldots, n\}$ such that $w_i \sqsubseteq v_{f(i)}$ for all $i \in \{1, \ldots, m\}$. The induced order on $W^*$ is known as the *monotone domination order*.

**Lemma 4.10** (Higman's Lemma [23]). *If $(W, \sqsubseteq)$ is a wqo then $(W^*, \sqsubseteq)$ is also a wqo.*

Next we use Higman's Lemma to construct a well-quasi-order on the state space of the transition system $\mathcal{W}_\mathcal{A}$. The first step is to define a class of *abstract configurations*, which are intended as canonical representatives of $\equiv$-equivalence classes of configurations.

**Definition 4.11.** An *abstract configuration* of the automaton $\mathcal{A}$ is a finite word over the alphabet $\Lambda$ of finite non-empty subsets of $S \times REG$, where $S$ is the set of locations of $\mathcal{A}$ and $REG$ is the set of regions.

Roughly speaking, each (concrete) $\mathcal{A}$-configuration $C$ gives rise to an abstract configuration as follows. First, $C$ is converted from a set to a list by ordering its elements according to the fractional part of their clock values. Then each clock value is replaced by the region it lies in. Formally, define an *abstraction function* $H : \wp(Q) \to \Lambda^*$, yielding an abstract configuration $H(C)$ for each configuration $C$ as follows. First, lift the function $reg$ to configurations by $reg(C) = \{(s, reg(v)) : (s, v) \in C\}$. Now given a configuration $C$, partition $C$ into a sequence of non-empty subsets $C_1, \ldots, C_n$, such that for all $(s, v) \in C_i$ and $(t, v') \in C_j$, $frac(v) \leq frac(v')$ iff $i \leq j$ (so $(s, v)$ and $(t, v')$ are in the same block $C_i$ iff $v$ and $v'$ have the same fractional part). Then define $H(C) = reg(C_1) reg(C_2) \ldots reg(C_n) \in \Lambda^*$.

**Example 4.12.** Consider the automaton $\mathcal{A}$ from Example 3.2. The maximum clock constant appearing in $\mathcal{A}$ is 1, thus the corresponding regions are $\mathsf{r}_0 = \{0\}$, $\mathsf{r}_1 = (0, 1)$, $\mathsf{r}_2 = \{1\}$ and $\mathsf{r}_3 = \{\top\}$. Given a concrete configuration $C = \{(s, 1), (t, 0.4), (s, 1.4), (t, 0.8)\}$, the corresponding abstract configuration $H(C)$ is the word $\{(s, \mathsf{r}_2), (s, \mathsf{r}_3)\} \{(t, \mathsf{r}_1)\} \{(t, \mathsf{r}_1)\}$.

The key fact about the abstraction function $H$, which is immediate from its definition, is that its kernel is the equivalence on configurations described in Definition 4.6:

**Proposition 4.13.** *Given $\mathcal{A}$-configurations $C$ and $D$, $C \equiv D$ iff $H(C) = H(D)$.*

Returning to Definition 4.11, notice that $\Lambda$, being finite, is trivially a wqo under the subset order. It follows from Lemma 4.10 that the set of abstract configurations is a wqo under the monotone domination order. Taking stock, we have defined a class of abstract configurations that is the quotient of the set of $\mathcal{A}$-configurations with respect to $\equiv$, and which carries a natural well-quasi-order. Next we show how to exploit this structure.

4.3. **Well-Structured Transition Systems.** The notion of *well-structured transition system (wsts)* provides a uniform framework for expressing decidability results about a variety of infinite-state systems, including Petri nets, broadcast protocols and lossy channel systems [1, 15]. Definition 4.14 presents a particular variant, called a *downward wsts* in [15].

**Definition 4.14.** A *well-structured transition system* is a triple $\mathcal{W} = (W, \preccurlyeq, \longrightarrow)$, where $(W, \longrightarrow)$ is a finitely-branching (unlabelled) transition system equipped with a wqo $\preccurlyeq$ such that:
- $\preccurlyeq$ is a decidable relation
- $Succ(w) := \{w' : w \longrightarrow w'\}$ is computable for each $w \in W$
- $\preccurlyeq$ is *downward-compatible*: if $w, v \in W$ with $w \preccurlyeq v$, then for any transition $v \longrightarrow v'$ there exists a matching sequence of transitions $w \ (\longrightarrow)^* \ w'$ with $w' \preccurlyeq v'$.

Note that downwards compatibility allows a single transition of $v$ to be matched by zero or more transitions of $w$.

**Theorem 4.15.** [15, Theorem 5.5] *Let $\mathcal{W} = (W, \preccurlyeq, \longrightarrow)$ be a wsts. Let $V \subseteq W$ be a downward-closed (i.e. $v' \preccurlyeq v$ and $v \in V$ imply $v' \in V$) decidable subset of $W$. Then, given a state $u \in W$, it is decidable whether there is a sequence of transitions starting at $u$ and ending in $V$.*

We now seek to apply Theorem 4.15 to the case at hand.

**Proposition 4.16.** *The transition system $\mathcal{W}_\mathcal{A}$ is a wsts (after forgetting the labels on the transitions).*

*Proof.* Define a quasi-order on the set of $\mathcal{A}$-configurations by $C \preccurlyeq D$ iff $H(C) \sqsubseteq H(D)$, i.e., the word $H(C)$ corresponding to $C$ is dominated by the word $H(D)$ corresponding to $D$. It is straightforward to see that $\preccurlyeq$ inherits the property of being a well-quasi-order from $\sqsubseteq$. Moreover $\preccurlyeq$ is a decidable relation on rational configurations, since $H$ is computable on rational configurations and $\sqsubseteq$ is decidable.

It remains to prove that $\preccurlyeq$ is downward compatible. Now suppose that $C \preccurlyeq D$ and that there is a transition $D \longrightarrow D'$. We show how to produce a matching sequence of transitions for $C$. To this end, it is helpful to first observe that $C \preccurlyeq D$ implies that there is a configuration $D_0 \subseteq D$ with $C \equiv D_0$. We now consider two cases according to whether the transition $D \longrightarrow D'$ arises from a $\Sigma$-labelled transition or an $\varepsilon$-labelled transition.

*$\Sigma$-labelled transitions.* Suppose that $D \xrightarrow{a} D'$ for some $a \in \Sigma$. Since $D_0 \subseteq D$ and since the successors of a configuration under discrete steps are computed pointwise (cf. Definition 3.4), there is a configuration $D'_0 \subseteq D'$ with $D_0 \xrightarrow{a} D'_0$. Now $C \equiv D_0$, so the Bisimulation Lemma yields a transition $C \xrightarrow{a} C'$ with $C' \equiv D'_0$. But $C' \equiv D'_0$ and $D'_0 \subseteq D'$ together imply $C' \preccurlyeq D'$.

$\varepsilon$-*transitions.* Suppose that $D \xrightarrow{\varepsilon} D'$. Then $D' = D + t$ for some $t \geq 0$, and, writing $D'_0 = D_0 + t$, we have $D'_0 \subseteq D'$. By Corollary 4.8 there exists a configuration $C'$ such that $C \xrightarrow{\varepsilon^*} C'$ and $C' \equiv D'_0$. But $C' \equiv D'_0$ and $D'_0 \subseteq D'$ together imply $C' \preccurlyeq D'$. □

We are now ready to state one of our main results.

**Theorem 4.17.** *Let $\mathcal{A}$ be a one-clock timed alternating automaton and let $\mathcal{B}$ be an Alur-Dill timed automaton. Then the language-emptiness problem '$L_f(\mathcal{A}) = \emptyset$?' and the language-inclusion problem '$L_f(\mathcal{B}) \subseteq L_f(\mathcal{A})$?' are both decidable.*

*Proof.* Since a configuration of $\mathcal{W}_\mathcal{A}$ is accepting if it only mentions accepting locations of $\mathcal{A}$, the set of accepting configurations of $\mathcal{W}_\mathcal{A}$ is downward-closed with respect to $\preccurlyeq$. By Proposition 4.15 it is decidable whether an accepting configuration of $\mathcal{W}_\mathcal{A}$ is reachable from the initial configuration. In turn this entails, by Proposition 4.9, that it is decidable whether an accepting configuration of $\mathcal{T}_\mathcal{A}$ is reachable from the initial configuration. But this question is equivalent to language emptiness for $\mathcal{A}$. This proves the first assertion of Theorem 4.17. The proof of the second assertion relies on the construction of a wsts representing the execution of $\mathcal{B}$ and $\mathcal{A}$ in parallel. We omit the details since we treat at length essentially the same construction in Section 8, where we consider a closely related language inclusion problem over infinite timed words. □

As noted earlier, these results have recently and independently been obtained by Lasota and Walukiewicz [25], also building on our earlier paper [28].

## 5. Metric Temporal Logic

In this section we define the syntax and semantics of Metric Temporal Logic (MTL). As discussed in the introduction, there are two different dense-time semantics for MTL: *event-based* and *state-based*, and for our concerns the difference is crucial. Following [16, 9, 10, 18, 19, 35], among others, we adopt an event-based semantics using timed words. A key observation about this semantics is that the temporal connectives quantify over a countable set of positions in a timed word. In contrast, the state-based semantics, adopted in, e.g., [7, 21, 32], associates a state to each point in real time, and the temporal connectives quantify over the whole time domain.[9] In the state-based semantics one can use a formula of the type $\Box(p \leftrightarrow \Diamond_{=1} q)$ to specify a perfect channel, whereas in the event-based semantics the same formula only specifies a channel with insertion errors (see Section 7). This observation helps understand why MTL is undecidable under the state-based semantics, whereas, at least over finite words, it is decidable in the event-based semantics (Theorem 6.5).

In the event-based semantics the atomic propositions in MTL refer to particular events, and the temporal connectives quantify over future events. This offers a natural idiom for reasoning about real-time behaviours, as we demonstrate in Example 5.4.

**Definition 5.1.** Given an alphabet $\Sigma$ of events, the formulas of MTL are built up from $\Sigma$ by Boolean connectives and time-constrained versions of the *next* operator $\bigcirc$ and the *until* operator $\mathcal{U}$ as follows:

$$\varphi ::= a \mid \textbf{true} \mid \varphi_1 \wedge \varphi_2 \mid \neg \varphi \mid \bigcirc_I \varphi \mid \varphi_1 \, \mathcal{U}_I \, \varphi_2 \, ,$$

---
[9]The state-based semantics views MTL as a subset of the monadic first-order theory of the non-negative reals, while the event-based semantics views MTL as a subset of a monadic first-order theory of the naturals with timestamps [9].

where $a \in \Sigma$, and $I \subseteq \mathbb{R}_{\geq 0}$ is an open, closed, or half-open interval with endpoints in $\mathbb{N} \cup \{\infty\}$. If $I = [0, \infty)$, then we omit the annotation $I$ in $\bigcirc_I$ and $\mathcal{U}_I$. We also use pseudo-arithmetic expressions to denote intervals. For example, the expression '$\geq 1$' denotes $[1, \infty)$ and '$= 1$' denotes the singleton $\{1\}$.

Additional temporal operators are defined following the usual conventions. We have the *constrained eventually* operator $\Diamond_I \varphi \equiv \mathbf{true}\, \mathcal{U}_I\, \varphi$, and the *constrained always* operator $\Box_I \varphi \equiv \neg \Diamond_I \neg \varphi$. We define a *dual until* operator via the standard duality: $\varphi_1\, \widetilde{\mathcal{U}}_I\, \varphi_2 \equiv \neg(\neg \varphi_1\, \mathcal{U}_I\, \neg \varphi_2)$. We also define the dual of the time-constrained next operator by $\widetilde{\bigcirc}_I \varphi \equiv \neg \bigcirc_I \neg \varphi$.[10]

**Definition 5.2.** Given a (finite or infinite) timed word $\rho = (\sigma, \tau)$ over alphabet $\Sigma$, a word position $i \leq |\rho|$, and an MTL formula $\varphi$, the satisfaction relation $(\rho, i) \models \varphi$ (read $\rho$ satisfies $\varphi$ at position $i$) is defined as follows:
- $(\rho, i) \models a$ iff $\sigma_i = a$
- $(\rho, i) \models \mathbf{true}$
- $(\rho, i) \models \varphi_1 \wedge \varphi_2$ iff $(\rho, i) \models \varphi_1$ and $(\rho, i) \models \varphi_2$
- $(\rho, i) \models \neg \varphi$ iff $(\rho, i) \not\models \varphi$
- $(\rho, i) \models \bigcirc_I \varphi$ iff $i < |\rho|$, $\tau_{i+1} - \tau_i \in I$ and $(\rho, i+1) \models \varphi$
- $(\rho, i) \models \varphi_1\, \mathcal{U}_I\, \varphi_2$ iff there exists $j$, $i \leq j \leq |\rho|$, such that $(\rho, j) \models \varphi_2$, $\tau_j - \tau_i \in I$, and $(\rho, k) \models \varphi_1$ for all $k$ with $i \leq k < j$.

For future reference it is also helpful to detail the semantics of the derived operators *dual until* and *dual next*:
- $(\rho, i) \models \widetilde{\bigcirc}_I \varphi$ iff $i = |\rho|$ or $\tau_{i+1} - \tau_i \notin I$ or $(\rho, i+1) \models \varphi$
- $(\rho, i) \models \varphi_1\, \widetilde{\mathcal{U}}_I\, \varphi_2$ iff for all $j$ such that $i \leq j \leq |\rho|$ and $\tau_j - \tau_i \in I$, either $(\rho, j) \models \varphi_2$ or there exists $k$ with $i \leq k < j$ and $(\rho, k) \models \varphi_1$.

We say that $\rho$ satisfies $\varphi$, denoted $\rho \models \varphi$, if $(\rho, 1) \models \varphi$. The set of finite models of an MTL formula $\varphi$ is given by $L_f(\varphi) = \{\rho \in T\Sigma^* : \rho \models \varphi\}$. The set of infinite models of $\varphi$ is given by $L_\omega(\varphi) = \{\rho \in T\Sigma^\omega : \rho \models \varphi\}$.

**Remark 5.3.** Note that in the semantics of MTL, time is measured relative to the occurrence of the first event of a timed word. In particular, the semantics is *translation invariant*: adding a fixed delay $d$ to each timestamp in a timed word does not change whether that word satisfies a formula or not. For this reason Wilke [35] restricts attention to timed words in which the first event has timestamp 0. In this case one can think of the first event as an initialisation event.

**Example 5.4.** The following example illustrates the convenience of event-based reasoning in the real-time setting. Consider a set of events $\Sigma = \{req_i, acq_i, rel_i : i = X, Y\}$ denoting the actions of two processes $X$ and $Y$ that request, acquire, and release a lock.
- $\Box(acq_X \rightarrow \Box_{<3} \neg acq_Y)$ says that $Y$ cannot acquire the lock less than 3 seconds after $X$ acquires the lock.
- $\Box(acq_X \rightarrow rel_X\, \widetilde{\mathcal{U}}_{<3}\, \neg acq_Y)$ says that $Y$ cannot acquire the lock less than 3 seconds after $X$ acquires the lock, unless $X$ first releases it.
- $\Box(req_X \rightarrow \Diamond_{<2}(acq_X \wedge \Diamond_{=1} rel_X))$ says that whenever $X$ requests the lock, it acquires the lock within 2 seconds and releases it exactly one second later.

---
[10] Note that, unlike $\bigcirc$ in LTL, $\bigcirc_I$ is not self-dual.

## 6. MTL over Finite Words

In this section we consider the *satisfiability problem* for MTL over finite words: 'Given an MTL formula $\varphi$, is $L_f(\varphi)$ nonempty?'. We also consider the following *model-checking problem*: 'Given an MTL formula $\varphi$ and an Alur-Dill timed automaton $\mathcal{B}$, is it the case that $L_f(\mathcal{B}) \subseteq L_f(\varphi)$?'. In both cases we show decidability by translating the MTL formulas into equivalent one-clock timed alternating automata and invoking Theorem 4.17. We also show that both problems have non-primitive recursive complexity.

6.1. **Translation to Automata.** By using disjunction, falsity, dual until and dual next, in addition to the standard MTL connectives, every formula can be put into a *negation normal form*, in which negation is only applied to events $a \in \Sigma$. Given an MTL formula $\varphi$ in negation normal form, we define a one-clock alternating automaton $\mathcal{A}_\varphi$ such that $L_f(\mathcal{A}_\varphi) = L_f(\varphi)$.

**Definition 6.1.** Define the *closure* of $\varphi$, denoted $cl(\varphi)$, as follows:
- $cl(\varphi)$ contains an element $\varphi_{init}$, called *the initial copy of* $\varphi$
- $cl(\varphi)$ contains each subformula of $\varphi$ whose outermost connective is $\mathcal{U}$ or $\widetilde{\mathcal{U}}$
- for each subformula $\bigcirc_I \psi$ of $\varphi$, $cl(\varphi)$ contains an element $(\bigcirc_I \psi)^r$, called the *residual copy* of $\bigcirc_I \psi$
- for each subformula $\widetilde{\bigcirc}_I \psi$ of $\varphi$, $cl(\varphi)$ contains an element $(\widetilde{\bigcirc}_I \psi)^r$, called the *residual copy* of $\widetilde{\bigcirc}_I \psi$.

The closure $cl(\varphi)$ forms the set of locations of $\mathcal{A}_\varphi$; thus states of $\mathcal{A}_\varphi$ are pairs $(\psi, v)$, where $\psi \in cl(\varphi)$ and $v$ is a clock value. We define the transition function $\delta$ so that the presence of state $(\psi, 0)$ in a configuration during a run of $\mathcal{A}_\varphi$ ensures that the input word satisfies $\psi$ at the current position. To enforce this requirement, when $\psi$ is encountered the automaton starts a fresh clock and thereafter propagates $\psi$ from configuration to configuration in the run until all the obligations that it stipulates are discharged.

**Definition 6.2.** Let $\varphi$ be an MTL formula in negation normal form. The automaton $\mathcal{A}_\varphi$ has set of locations $cl(\varphi)$. The initial location is $\varphi_{init}$ and the accepting locations are those elements of $cl(\varphi)$ of the form $\varphi_1 \, \widetilde{\mathcal{U}}_I \, \varphi_2$ or $(\widetilde{\bigcirc}_I \psi)^r$. In order to give a smooth recursive definition of the transition function $\delta$, we define $\delta(\psi, a)$ for all subformulas $\psi$ of $\varphi$, not just those in $cl(\varphi)$. The definition is given by the following clauses, where $a, b \in \Sigma$:

$$\begin{aligned}
\delta(\varphi_{init}, a) &= x.\delta(\varphi, a) \\
\delta(\psi_1 \vee \psi_2, a) &= \delta(\psi_1, a) \vee \delta(\psi_2, a) \\
\delta(\psi_1 \wedge \psi_2, a) &= \delta(\psi_1, a) \wedge \delta(\psi_2, a) \\
\delta(\psi_1 \, \mathcal{U}_I \, \psi_2, a) &= ((x.\delta(\psi_2, a)) \wedge x \in I) \vee ((x.\delta(\psi_1, a)) \wedge (\psi_1 \, \mathcal{U}_I \, \psi_2)) \\
\delta(\psi_1 \, \widetilde{\mathcal{U}}_I \, \psi_2, a) &= ((x.\delta(\psi_2, a)) \vee x \notin I) \wedge ((x.\delta(\psi_1, a)) \vee (\psi_1 \, \widetilde{\mathcal{U}}_I \, \psi_2)) \\
\delta(\bigcirc_I \psi, a) &= x.(\bigcirc_I \psi)^r \\
\delta((\bigcirc_I \psi)^r, a) &= (x \in I) \wedge x.\delta(\psi, a) \\
\delta(\widetilde{\bigcirc}_I \psi, a) &= x.(\widetilde{\bigcirc}_I \psi)^r \\
\delta((\widetilde{\bigcirc}_I \psi)^r, a) &= (x \notin I) \vee x.\delta(\psi, a)
\end{aligned}$$

$$\delta(b, a) = \begin{cases} \textbf{true} & \text{if } a = b \\ \textbf{false} & \text{if } a \neq b \end{cases}$$

$$\delta(\neg b, a) = \begin{cases} \textbf{false} & \text{if } a = b \\ \textbf{true} & \text{if } a \neq b. \end{cases}$$

**Remark 6.3.** Notice the connection between the notion of duality for MTL formulas and the notion of duality for transition functions (described in Subsection 3.2). In particular, we have $\overline{\delta(\psi_1 \, \mathcal{U}_I \, \psi_2, a)} = \delta(\psi_1 \, \widetilde{\mathcal{U}}_I \, \psi_2, a)$ and $\overline{\delta((\bigcirc_I \psi)^r, a)} = \delta((\widetilde{\bigcirc}_I \psi)^r, a)$.

**Proposition 6.4.** *Given an MTL formula $\varphi$ in negation normal form, $L_f(\mathcal{A}_\varphi) = L_f(\varphi)$.*

*Proof.* We first show that $L_f(\mathcal{A}_\varphi) \subseteq L_f(\varphi)$. To this end, let $\rho = (\sigma, \tau)$ be a timed word in $L_f(\mathcal{A}_\varphi)$, with $|\rho| = n$. As usual, write $d_i = \tau_i - \tau_{i-1}$ for $1 \leq i \leq n$. Suppose that $\mathcal{A}_\varphi$ has an accepting run on $\rho$:

$$C_0 \xrightarrow{d_1} C_1 \xrightarrow{\sigma_1} C_2 \xrightarrow{d_2} C_3 \xrightarrow{\sigma_2} \cdots \xrightarrow{\sigma_n} C_{2n}.$$

We claim that for each subformula $\psi$ of $\varphi$ and each $i$ such that $1 \leq i \leq n$, $(\rho, i) \models \psi$ whenever $C_{2i} \models_0 \delta(\psi, \sigma_i)$. We prove this claim by structural induction on $\psi$.

The base case, in which $\psi \equiv a$ or $\psi \equiv \neg a$ for an atomic formula $a \in \Sigma$, is immediate. The only non-trivial cases in the induction step are when the outermost connective of $\psi$ is a temporal modality. We consider the cases $\psi \equiv \bigcirc_I \psi_1$ and $\psi \equiv \psi_1 \, \mathcal{U}_I \, \psi_2$; the cases for the dual temporal connectives are similar.

**Case** $\psi \equiv \bigcirc_I \psi_1$. If $C_{2i} \models_0 \delta(\psi, \sigma_i)$ then, since $\delta(\psi, \sigma_i) = x.(\bigcirc_I \psi_1)^r$, we must have $((\bigcirc_I \psi_1)^r, 0) \in C_{2i}$. In turn, this entails that $C_{2i+2} \models_0 \delta(\psi_1, \sigma_{i+1})$ and $\tau_{i+1} - \tau_i \in I$. By the induction hypothesis we have $(\rho, i+1) \models \psi_1$, whence $(\rho, i) \models \bigcirc_I \psi_1$.

**Case** $\psi \equiv \psi_1 \, \mathcal{U}_I \, \psi_2$. Suppose $C_{2i} \models_0 \delta(\psi, \sigma_i)$. We consider two possibilities, corresponding to the two disjuncts in the definition of $\delta(\psi, \sigma_i)$. One possibility is that $C_{2i} \models_0 \delta(\psi_2, \sigma_i)$ and $0 \in I$. In this case, by the induction hypothesis, we have $(\rho, i) \models \psi_2$, whence $(\rho, i) \models \psi_1 \, \mathcal{U}_I \, \psi_2$. On the other hand, we may have $C_{2i} \models_0 \delta(\psi_1, \sigma_i)$ and $(\psi, 0) \in C_{2i}$. Then the definition of the transition function $\delta$ ensures that for each successive value of $j \geq i$ we have that $C_{2j} \models \delta(\psi_1, \sigma_j)$ and $(\psi, \tau_j - \tau_i) \in C_{2j}$ until at some point $C_{2j} \models \delta(\psi_2, \sigma_j)$ and $\tau_j - \tau_i \in I$. (Note that the latter must eventually occur since $\psi$ is not an accepting location.) From the induction hypothesis it is clear that this implies that $(\rho, i) \models \psi$. This completes the proof of the claim.

Having proved the claim, we observe that $(\varphi_{init}, 0) \in C_0$, and, since $\delta(\varphi_{init}, \sigma_1) = x.\delta(\varphi, \sigma_1)$, we have $C_2 \models_0 \delta(\varphi, \sigma_1)$. Thus, applying the claim in case $i = 1$ and $\psi \equiv \varphi$, we immediately get that $\rho \models \varphi$. This completes the proof that $L_f(\mathcal{A}_\varphi) \subseteq L_f(\varphi)$.

It remains to show the converse inclusion: $L_f(\varphi) \subseteq L_f(\mathcal{A}_\varphi)$. To this end, we show that, up to renaming of locations, $\mathcal{A}_{\neg\varphi} = (\mathcal{A}_\varphi)^c$, that is, the automaton representing $\neg\varphi$ is the complement of the automaton representing $\varphi$. Indeed the set of locations of $(\mathcal{A}_\varphi)^c$ is the same as the set of locations of $\mathcal{A}_\varphi$: it is just $cl(\varphi)$. On the other hand, the set of locations of $\mathcal{A}_{\neg\varphi}$ is $cl(\neg\varphi)$, which consists of the duals of the formulas in $cl(\varphi)$. Thus the map sending a formula to its dual is a bijection between the locations of $\mathcal{A}_{\neg\varphi}$ and $(\mathcal{A}_\varphi)^c$. But now Remark 6.3 shows that the respective transition functions of $\mathcal{A}_{\neg\varphi}$ and $(\mathcal{A}_\varphi)^c$ are identical (modulo the bijection between the respective sets of locations).

Now, using the inclusion that we have just proved, we have

$$T\Sigma^* \setminus L_f(\mathcal{A}_\varphi) = L_f((\mathcal{A}_\varphi)^c) = L_f(\mathcal{A}_{\neg\varphi}) \subseteq L_f(\neg\varphi) = T\Sigma^* \setminus L_f(\varphi).$$

But this directly gives $L_f(\varphi) \subseteq L_f(\mathcal{A}_\varphi)$, which completes the proof. □

In conjunction with Theorem 4.17, Proposition 6.4 immediately yields:

**Theorem 6.5.** *The satisfiability and the model-checking problems for MTL over finite words are both decidable.*

## 7. Complexity

In this section, using a result of Schnoebelen [33] about lossy channel systems, we prove that the satisfiability and model-checking problems for MTL have non-primitive recursive complexity.

A *channel machine* consists of a finite-state automaton acting on an unbounded FIFO channel, or queue. More precisely, a channel machine is a tuple $\mathcal{C} = (S, M, \Delta)$, where $S$ is a finite set of *control states*, $M$ is a finite set of *messages*, and $\Delta \subseteq S \times \Sigma \times S$ is the transition relation over label set $\Sigma = \{m!, m? : m \in M\}$. A transition labelled $m!$ writes message $m$ to the tail of the channel, and a transition labelled $m?$ reads message $m$ from the head of the channel.

We define an operational semantics for channel machines as follows. A *global state* of $\mathcal{C}$ is a pair $\gamma = (s, x)$, where $s \in S$ is the control state and $x \in M^*$ represents the contents of the channel. The rules in $\Delta$ induce a $\Sigma$-labelled transition relation on the set of global states thus: $(s, m!, t) \in \Delta$ yields a transition $(s, x) \xrightarrow{m!} (t, x \cdot m)$ that writes $m \in M$ to the tail of the channel, and $(s, m?, t) \in \Delta$ yields a transition $(s, m \cdot x) \xrightarrow{m?} (t, x)$ that reads $m \in M$ from the head of the channel. If we only allow the transitions indicated above, then we call $\mathcal{C}$ an *error-free* channel machine. A *computation* of such a machine is a finite sequence of transitions between global states

$$(s_0, x_0) \xrightarrow{\alpha_0} (s_1, x_1) \xrightarrow{\alpha_1} \cdots \xrightarrow{\alpha_{n-1}} (s_n, x_n). \tag{7.1}$$

We also consider channel machines that are subject to *insertion errors*. Given $x, y \in M^*$, write $x \sqsubseteq y$ if $x$ is a subword of $y$, i.e., $x$ can be obtained from $y$ by deleting any number of letters; for example, sub $\sqsubseteq$ st<u>ub</u>born, as indicated by the underlining. (This is a special instance of the monotone domination order introduced earlier.) Following [33] we model *insertion errors* by extending the transition relation on global states with the following clause: if $(s, x) \xrightarrow{\alpha} (t, y)$, $x' \sqsubseteq x$ and $y \sqsubseteq y'$, then $(s, x') \xrightarrow{\alpha} (t, y')$. Dually, we define *lossy channel machines* by adding a clause: if $(s, x) \xrightarrow{\alpha} (t, y)$, $x \sqsubseteq x'$ and $y' \sqsubseteq y$, then $(s, x') \xrightarrow{\alpha} (t, y')$. The notion of a computation of a channel machine with insertion errors or lossiness errors is defined analogously to the error-free case, but with the extended transition relations.

The *control-state reachability problem* asks, given a channel machine $\mathcal{C} = (S, M, \Delta)$ and two distinct control states $s_{init}, s_{fin} \in S$, whether there is a finite computation of $\mathcal{C}$ starting in global state $(s_{init}, \varepsilon)$ and ending in global state $(s_{fin}, x)$ for some $x \in M^*$. This problem was proved to be decidable for lossy channel machines by Abdulla and Jonsson [4]. Later Schnoebelen [33] showed that it has non-primitive recursive complexity.

The *dual control-state reachability problem* asks, given a channel machine $\mathcal{C} = (S, M, \Delta)$ and two distinct control states $s_{init}, s_{fin} \in S$, whether there is a finite computation of $\mathcal{C}$ starting in control state $(s_{init}, x)$ and ending in state $(s_{fin}, \varepsilon)$, for some initial channel contents $x \in M^*$.

Note that the difference between the control-state reachability problem and the dual control-state reachability problem depends on whether the initial or final channel is required to be empty. This difference is significant. For instance, the control-state reachability problem is trivial for channel machines with insertion errors. In this case there is a computation from $(s_{init}, \varepsilon)$ to $(s_{fin}, x)$ for some $x \in M^*$ iff there is a path from $s_{init}$ to $s_{fin}$ in the underlying control automaton. Indeed, given such a path we can always construct a matching computation of the channel machine by using insertion errors to ensure that every read-transition along the path is enabled. In contrast, for the dual control-state reachability problem we have the following result.

**Proposition 7.1.** *The dual control-state reachability problem for channel machines with insertion errors has non-primitive recursive complexity.*

*Proof.* Given a channel machine $\mathcal{C} = (S, M, \Delta)$, the *opposite channel machine* is defined by $\mathcal{C}^{\mathrm{op}} = (S, M, \Delta^{\mathrm{op}})$ where
$$\Delta^{\mathrm{op}} = \{(s, m!, t) : (t, m?, s) \in \Delta\} \cup \{(s, m?, t) : (t, m!, s) \in \Delta\}.$$
Note that $\mathcal{C}$ has a computation from $(s, x)$ to $(t, y)$ with lossiness errors iff $\mathcal{C}^{\mathrm{op}}$ has a computation from $(t, y^{\mathrm{op}})$ to $(s, x^{\mathrm{op}})$ with insertion errors, where $(-)^{\mathrm{op}} \colon M^* \to M^*$ reverses the order of a word. Thus the dual control-state reachability problem for $\mathcal{C}$ with insertion errors is equivalent to the control-state reachability problem for $\mathcal{C}^{\mathrm{op}}$ with lossiness errors. But, as we mentioned above, this last problem is known to be decidable with non-primitive recursive complexity. □

**Theorem 7.2.** *The satisfiability and model-checking problems for MTL over finite words have non-primitive recursive complexity.*

*Proof.* We give a reduction of the dual control-state reachability problem for channel machines with insertion errors to the satisfiability problem for MTL. Let $\mathcal{C} = (S, M, \Delta)$ and $s_{init}, s_{fin} \in S$ be an instance of the dual control-state reachability problem. The idea is to encode computations of $\mathcal{C}$ as timed words over the alphabet $\Sigma = S \cup \{m!, m? : m \in M\}$. For instance, the computation (7.1) is represented by a timed word whose sequence of events is $s_0\alpha_0 s_1 \ldots \alpha_{n-1} s_n$. In this encoding the key idea is to choose timestamps that mirror the FIFO discipline of the channel. This is done by requiring that every write-event $m!$ be followed one time unit later by a matching read-event $m?$.

In the following we describe an MTL formula $\varphi_{REACH}$ that describes all timed words that encode computations of $\mathcal{C}$ starting in $s_{init}$ and ending in state $s_{fin}$ with empty channel. Thus $\varphi_{REACH}$ is satisfiable iff $\mathcal{C}$ is a positive instance of the dual control-state reachability problem.

We use the formula $\varphi_{CHAN}$, below, to capture the behaviour of a channel: every write-event is followed one time unit later by a matching read-event. However, there is no guarantee that every read-event is *preceded* one time unit earlier by a write-event, so the channel may have insertion errors.
$$\varphi_{CHAN} \equiv \bigwedge\nolimits_{m \in M} \Box(m! \to \Diamond_{=1} m?).$$

In order that there be no confusion in matching write-events with their corresponding subsequent read-events, we require that time be strongly monotonic (no two events can occur at the same time). This is captured by the formula $\varphi_{SM}$:
$$\varphi_{SM} \equiv (\bigcirc_{>0} \mathbf{true}) \, \mathcal{U} \, \neg \bigcirc \mathbf{true}.$$

We encode the finite control of $\mathcal{C}$ using the formula $\varphi_{CONT}$:

$$\varphi_{CONT} \equiv \bigwedge_{s \in S} (s \rightarrow \bigvee_{(s,\mu,t) \in \Delta} (\bigcirc \mu \wedge \bigcirc \bigcirc t)).$$

We then use $\varphi_{RUN}$ to assert that a run must start in control state $s_{init}$ and obey the discrete controller until it terminates in control state $s_{fin}$ with empty channel:

$$\varphi_{RUN} \equiv s_{init} \wedge (\varphi_{CONT} \,\mathcal{U}\, (s_{fin} \wedge \neg \bigcirc \mathbf{true})).$$

Finally, we combine all these requirements into $\varphi_{REACH}$:

$$\varphi_{REACH} \equiv \varphi_{CHAN} \wedge \varphi_{SM} \wedge \varphi_{RUN}.$$

Suppose we are given a timed word $\rho$ satisfying $\varphi_{REACH}$; then we can construct a computation of $\mathcal{C}$ as follows. First, observe that $\rho$ consists of an alternating sequence of events from $S$ and events from $\{m!, m? : m \in M\}$. This gives the sequence of control states and transitions in the desired computation; it remains to construct the contents of the channel at each control state. Suppose event $s \in S$ occurs at some point along $\rho$ with timestamp $t$. Then the channel contents associated to this occurrence of $s$ is the sequence of read-events occurring in $\rho$ in the time interval $(t, t+1)$. Observe how this definition ensures that a message can only be read from the head of the channel, and how each write-event adds a message to the tail of the channel. Finally, observe that any timed word satisfying $\varphi_{REACH}$ must have $s_{fin}$ as its last event; this ensures that the channel is empty at that point.

Conversely, suppose we are given a computation of $\mathcal{C}$,

$$(s_0, x_0) \xrightarrow{\alpha_0} (s_1, x_1) \xrightarrow{\alpha_1} \cdots \xrightarrow{\alpha_{n-1}} (s_n, x_n)$$

with $s_0 = s_{init}$, $s_n = s_{fin}$ and $x_n = \varepsilon$. We then derive a timed word $\rho = (\sigma, \tau)$ that satisfies $\varphi_{REACH}$. We define $\sigma = s_0 \alpha_0 s_1 \alpha_1 \ldots s_n$; this guarantees that $\rho$ satisfies $\varphi_{RUN}$. It remains to choose a sequence of timestamps $\tau$ such that $\varphi_{CHAN} \wedge \varphi_{SM}$ is also satisfied.

Since the given computation of $\mathcal{C}$ ends with the empty channel, every message that is written to the channel is eventually read from the channel. Thus for each write-event $m!$ in $\sigma$ there is a 'matching' read-event $m?$ later on. We choose the sequence of timestamps $\tau$ so that each such matching pair is separated by one time unit. This captures the FIFO discipline of the channel: messages are read from the channel in the same order that they were written to the channel. Formally we choose the $\tau_i$ sequentially, starting with $\tau_1 = 0$ and maintaining the following invariant: $\tau_i$ is chosen such that for each matching pair $\sigma_j = m!$ and $\sigma_k = m?$, if $j < k = i$ then $\tau_i - \tau_j = 1$, and if $j < i < k$ then $\tau_i - \tau_j < 1$. It is clearly possible to do this using the density of time.

Thus a channel machine $\mathcal{C} = (S, M, \Delta)$ and pair of control states $s_{init}, s_{fin} \in S$ is a positive instance of the dual reachability problem iff the formula $\varphi_{REACH}$ is satisfiable. This shows that the satisfiability problem for MTL has non-primitive recursive complexity.

Finally, consider a universal Alur-Dill timed automaton, i.e., one that accepts all finite timed words. Model checking this automaton against a given MTL formula is equivalent to asking whether the formula is valid, i.e., whether its negation is unsatisfiable. The complexity of model checking MTL is therefore also non-primitive recursive. □

## 8. Infinite Words: Safety MTL

In this section we adapt constructions from Section 4 to prove the decidability of the model-checking problem over infinite words for a subset of MTL, called *Safety MTL*. Safety MTL consists of those MTL formulas whose negation normal form only includes instances of the constrained until operator $\mathcal{U}_I$ in which interval $I$ has bounded length. Note that no restrictions are placed on the dual-until operator $\widetilde{\mathcal{U}}_I$.

Safety MTL can express time-bounded response properties, but not arbitrary response formulas. For instance, the formulas $\varphi_1 \equiv \Box(a \to \Diamond_{=1} b)$ and $\varphi_2 \equiv \Box(a \to \Diamond_{\leq 5}(b \wedge \Diamond_{=1} c))$ are in Safety MTL, but $\varphi_3 \equiv \Diamond a$ is not. Note in passing that, intuitively, $\varphi_2$ is much harder to model check than $\varphi_1$. To find a counterexample to $\varphi_1$, one need only guess an $a$-event, and check that there is no $b$-event one time unit later—a task requiring only one clock. On the other hand, to find a counterexample to $\varphi_2$ one must not only guess an $a$-event, but also check that every $b$-event in the ensuing five time units fails to have a matching $c$-event one time unit later—a task requiring a potentially unbounded number of clocks.

To explain the name Safety MTL, recall from [17] that a language $L \subseteq T\Sigma^\omega$ defines a *safety property relative to the divergence of time* if for every $\rho \notin L$ there exists $n \in \mathbb{N}$ such that no infinite timed word in $T\Sigma^\omega$ extending $\rho[1\ldots n]$ is contained in $L$. In this case we say that $\rho[1\ldots n]$ is a *bad prefix* of $\rho$.

**Proposition 8.1.** *For every Safety MTL formula $\varphi$, $L_\omega(\varphi)$ is a safety property relative to the divergence of time.*

*Proof.* It is straightforward to prove this result by structural induction on $\varphi$. However, we do not give details since we do not use this result in the sequel and since, in any case, it follows directly from Proposition 8.2 and Proposition 8.3. □

To model check a Safety MTL formula $\varphi$ on an Alur-Dill automaton $\mathcal{B}$ we need only check whether any of the bad prefixes of $\varphi$ are prefixes of words accepted by $\mathcal{B}$. We can do this by invoking a variant of the idea used in the proof of Theorem 4.17. To set up this model-checking procedure we first define a translation of $\varphi$ into a one-clock alternating automaton $\mathcal{A}_\varphi^{safe}$ in which every location is accepting.

$\mathcal{A}_\varphi^{safe}$ is a modification of the automaton $\mathcal{A}_\varphi$ from Section 6.1. $\mathcal{A}_\varphi^{safe}$ has the same alphabet, locations and initial location as $\mathcal{A}_\varphi$, but we declare every location of $\mathcal{A}_\varphi^{safe}$ to be accepting. To compensate for this last change, we modify a single clause in the definition of the transition function $\delta$—the clause for $\varphi_1 \mathcal{U}_I \varphi_2$—as indicated below.

$$\delta(\varphi_1\,\mathcal{U}_I\,\varphi_2, a) = ((x.\delta(\varphi_2, a)) \wedge x \in I) \vee$$
$$((x.\delta(\varphi_1, a)) \wedge (\varphi_1\,\mathcal{U}_I\,\varphi_2) \wedge (x \leq \sup(I)))\,.$$

Intuitively, the above definition uses a 'timeout' rather than an acceptance condition to ensure that the second argument of $\mathcal{U}_I$ eventually becomes true. In a non-Zeno run, the automaton cannot get stuck forever in location $\varphi_1\,\mathcal{U}_I\,\varphi_2$ since the clock constraints in the definition of $\delta(\varphi_1\,\mathcal{U}_I\,\varphi_2, a)$ only allow transitions when the value of clock $x$ is no greater than $\sup(I)$.

Recall that so far we have only considered alternating automata on finite words. In order to state the correctness of the definition of $\mathcal{A}_\varphi^{safe}$ we consider runs of timed alternating automata on infinite words. Our task is simplified by the fact that we only consider automata in which every location is accepting. (Technically this means that, as with automata over finite words, we can elide the tree structure that is usually associated with runs of alternating automata.) Suppose then that $\mathcal{A}$ is a timed alternating automaton in which every location is accepting. A run of $\mathcal{A}$ on an infinite timed word $\rho = (\sigma, \tau)$ is an infinite alternating sequence of time delays and discrete steps in $\mathcal{T}_\mathcal{A}$:

$$C_0 \overset{d_1}{\rightsquigarrow} C_1 \overset{\sigma_1}{\longrightarrow} C_2 \overset{d_2}{\rightsquigarrow} C_3 \overset{\sigma_2}{\longrightarrow} \cdots \overset{d_n}{\rightsquigarrow} C_{2n-1} \overset{\sigma_n}{\longrightarrow} \cdots,$$

where $C_0$ is the initial configuration and $d_i = \tau_i - \tau_{i-1}$. We define $L_\omega(\mathcal{A})$ to be the set of non-Zeno timed words $\rho \in T\Sigma^\omega$ over which $\mathcal{A}$ has a run. (Since every location of $\mathcal{A}$ is accepting, there is no need to consider an acceptance condition here.)

**Proposition 8.2.** $L_\omega(\varphi) = L_\omega(\mathcal{A}_\varphi^{safe})$ *for each Safety MTL formula* $\varphi$.

*Proof.* The proof of Proposition 6.4 carries over almost verbatim to the present setting. Referring to the details of that proof, the only change is to observe that it is the 'timeout' in the definition of $\delta(\varphi_1 \, \mathcal{U}_I \, \varphi_2, a)$, rather than the fact that $\varphi_1 \, \mathcal{U}_I \, \varphi_2$ is non-accepting, that ensures that whenever $(\varphi_1 \, \mathcal{U}_I \, \varphi_2, 0)$ lies in some configuration $C_{2i}$ in a run, then there exists $j \geq i$ such that $C_{2j} \models \delta(\varphi_2, \sigma_j)$. □

8.1. **The Model-Checking Procedure.** In this section, let $\mathcal{B}$ be an Alur-Dill timed automaton with $n$ clocks, and let $\mathcal{A}$ be a one-clock alternating automaton in which every location is accepting. We describe a decision procedure for the model-checking problem '$L_\omega(\mathcal{B}) \subseteq L_\omega(\mathcal{A})$?'. Combining this procedure with Proposition 8.2 gives a method for model checking Safety MTL formulas on Alur-Dill automata.

The following proposition enables us to decide whether $L_\omega(\mathcal{B}) \subseteq L_\omega(\mathcal{A})$, while only considering finite runs of $\mathcal{A}$. The idea is that for any word $\rho \in T\Sigma^\omega \setminus L_\omega(\mathcal{A})$, there is a finite bad prefix $\rho[1 \ldots n]$ none of whose (non-Zeno) extensions lies in $L_\omega(\mathcal{A})$.

**Proposition 8.3.** *Let* $\mathcal{A}$ *be a timed alternating automaton in which every state is accepting. Then* $\rho \in T\Sigma^\omega \setminus L_\omega(\mathcal{A})$ *iff there exists* $n \in \mathbb{N}$ *such that* $\rho[1 \ldots n] \in L_f(\mathcal{A}^c)$.

*Proof.* We first consider the 'if' direction. Suppose that $\rho[1 \ldots n] \in L_f(\mathcal{A}^c)$.[11] By Proposition 3.7 there can be no run of $\mathcal{A}$ on the finite word $\rho[1 \ldots n]$. (Any such run would be accepting, since every location of $\mathcal{A}$ is accepting.) *A fortiori* there can be no run of $\mathcal{A}$ on $\rho$.

Now we show the 'only if' direction. If $\rho \notin L_\omega(\mathcal{A})$ then $\mathcal{A}$ does not have a run on $\rho$. Moreover we observe that for each $n \geq 1$ there are only finitely many ways to extend a run of $\mathcal{A}$ on the finite prefix $\rho[1 \ldots n]$ to a run on $\rho[1 \ldots (n+1)]$. Thus, by König's lemma, there exists $n \in \mathbb{N}$ such that $\mathcal{A}$ does not have a run on $\rho[1 \ldots n]$. For this choice of $n$ the complement automaton $\mathcal{A}^c$ accepts $\rho[1 \ldots n]$. □

From this point on, the explanation of the model-checking procedure closely follows Section 4. In fact, the remainder of this section recapitulates definitions and propositions from Section 4, *mutatis mutandis*. Briefly, the main difference between Section 4 and the present section is that rather than just considering a wsts generated by a timed alternating automaton, we consider a wsts generated by the timed alternating automaton $\mathcal{A}^c$ and the Alur-Dill automaton $\mathcal{B}$ executing in parallel. We reduce the language emptiness problem '$L_\omega(\mathcal{B}) \cap L_\omega(\mathcal{A}^c) = \emptyset$?' (which is equivalent to '$L_\omega(\mathcal{B}) \subseteq L_\omega(\mathcal{A})$?') to reachability on this wsts.

---

[11]Note that since none of the locations of $\mathcal{A}^c$ is accepting, $\mathcal{A}^c$ can only accept a word by moving to the empty configuration.

Denote by $c_{\max}$ the maximum clock constant appearing in $\mathcal{A}$ and $\mathcal{B}$, and let $\mathsf{Val} = [0, c_{\max}] \cup \{\top\}$ be a set of clock values appropriate to $\mathcal{A}$ and $\mathcal{B}$. Recall that a state of $\mathcal{B}$ is a pair $\gamma = (s, \mathbf{v})$, where $s$ is a location of $\mathcal{B}$ and $\mathbf{v} \in \mathsf{Val}^n$ is a clock valuation. Define a $\mathcal{B}$-$\mathcal{A}^c$-*configuration* to be a pair $(\gamma, C)$, where $\gamma$ is a state of $\mathcal{B}$ and $C$ is a configuration of $\mathcal{A}^c$. Following the pattern of Definition 4.4, we define a labelled transition system $\mathcal{T}_{\mathcal{B}, \mathcal{A}^c}$, representing $\mathcal{B}$ and $\mathcal{A}^c$ executing in parallel.

**Definition 8.4.** The set of states of $\mathcal{T}_{\mathcal{B}, \mathcal{A}^c}$ is the set of $\mathcal{B}$-$\mathcal{A}^c$-configurations. Following Definition 4.4 we define an ($\mathbb{R}_{\geq 0}$)-labelled delay-step transition relation by $(\gamma, C) \overset{t}{\rightsquigarrow} (\gamma + t, C + t)$ for $t \geq 0$, and a $\Sigma$-labelled discrete-step transition relation by $(\gamma, C) \xrightarrow{a} (\gamma', C')$ if $\gamma \xrightarrow{a} \gamma'$ in $\mathcal{T}_\mathcal{B}$ and $C \xrightarrow{a} C'$ in $\mathcal{T}_{\mathcal{A}^c}$, where $a \in \Sigma$.

A configuration $(\gamma, C)$ of $\mathcal{T}_{\mathcal{B}, \mathcal{A}^c}$ is said to be *initial* if $\gamma$ is the initial state of $\mathcal{B}$ and $C$ is the initial configuration of $\mathcal{A}^c$. Recall that $\mathcal{A}^c$ can only accept a word by moving to the empty configuration. Thus a timed word $\rho \in L_\omega(\mathcal{B})$ fails to lie in $L_\omega(\mathcal{A})$ iff there is a computation of $\mathcal{A}^c$ on a finite prefix of $\rho$ that reaches $\emptyset$. Motivated by this observation, we say that a $\mathcal{B}$-$\mathcal{A}^c$-configuration $(\gamma, C)$ is *doomed* if $C = \emptyset$ (i.e., $\mathcal{A}^c$ has reached an accepting configuration) and $\mathcal{B}$ can accept some infinite non-Zeno word starting in state $\gamma$. Then $L_\omega(\mathcal{B}) \not\subseteq L_\omega(\mathcal{A})$ iff there is a doomed configuration $(\gamma, \emptyset)$ that is reachable from the initial configuration of $\mathcal{T}_{\mathcal{B}, \mathcal{A}^c}$. Below we sketch how we can use Theorem 4.15 to prove that this reachability problem is decidable.

To set up the application of Theorem 4.15 we reuse constructions from Section 4 to show that $\mathcal{T}_{\mathcal{B}, \mathcal{A}^c}$ contains a sub-transition-system $\mathcal{W}_{\mathcal{B}, \mathcal{A}^c}$ that is a wsts. The first step is to adapt the notion of the time successor of a configuration to the present setting.

**Definition 8.5.** Let $(\gamma, C)$ be a $\mathcal{B}$-$\mathcal{A}^c$-configuration, where $\gamma = (s, \mathbf{v})$, and let $E = \{v_i : 1 \leq i \leq n\} \cup \{v : (t, v) \in C\}$ be the set of clock values appearing in $(\gamma, C)$. Write $\mu = \max\{frac(v) : v \in E\}$ for the maximum fractional part of the clock values in $E$. Now define the *time successor* of $(\gamma, C)$ to be the configuration $next(\gamma, C) = (\gamma + d, C + d)$, where $d = (1 - \mu)/2$ if $E$ contains an integer, and $d = 1 - \mu$ otherwise.

**Definition 8.6.** Define the labelled transition system $\mathcal{W}_{\mathcal{B}, \mathcal{A}^c}$ as follows.
- **Alphabet.** The alphabet of $\mathcal{W}_{\mathcal{B}, \mathcal{A}^c}$ is $\Sigma \cup \{\varepsilon\}$.
- **States.** The states of $\mathcal{W}_{\mathcal{B}, \mathcal{A}^c}$ are those configurations $(\gamma, C)$ in which all clock values are rational (henceforth call such configurations rational).
- **Transitions.** Each configuration $(\gamma, C)$ makes a unique $\varepsilon$-transition to its time successor $next(\gamma, C)$. For $a \in \Sigma$, we declare that $(\gamma, C) \xrightarrow{a} (\gamma', C')$ in $\mathcal{W}_{\mathcal{B}, \mathcal{A}^c}$ iff $(\gamma, C) \xrightarrow{a} (\gamma', C')$ in $\mathcal{T}_{\mathcal{B}, \mathcal{A}^c}$.

Continuing to shadow the development in Section 4, we adapt the Bisimulation Lemma, Lemma 4.7, to the present setting. We define an equivalence relation $\equiv$ on $\mathcal{B}$-$\mathcal{A}^c$ configurations that abstracts away from precise clock values, recording only their integer parts and the relative order of their fractional parts.

**Definition 8.7.** Suppose that $(\gamma, C)$ and $(\gamma', C')$ are $\mathcal{B}$-$\mathcal{A}^c$ configurations such that $\gamma = (s, (v_1, \ldots, v_n))$, $\gamma' = (s', (v'_1, \ldots, v'_n))$, $C = \{(s_i, u_i)\}_{i \in I}$ and $C' = \{(s'_i, u'_i)\}_{i \in I}$. (In particular, we require that $C$ and $C'$ have the same cardinality.) Then we define $(\gamma, C) \equiv (\gamma', C')$ if the following hold, where $\bowtie \in \{<, =, >\}$:
- $s = s'$ and $s_i = s'_i$ for each $i \in I$

- $u_i \sim u'_i$ for $i \in I$ and $v_j \sim v'_j$ for $j \in \{1, \ldots, n\}$
- $frac(u_i) \bowtie frac(u_j)$ iff $frac(u'_i) \bowtie frac(u'_j)$ for $i, j \in I$
- $frac(v_i) \bowtie frac(v_j)$ iff $frac(v'_i) \bowtie frac(v'_j)$ for $i, j \in \{1, \ldots, n\}$
- $frac(u_i) \bowtie frac(v_j)$ iff $frac(u'_i) \bowtie frac(v'_j)$ for $i \in I$, $j \in \{1, \ldots, n\}$.

The first four clauses of this definition ensure that $(\gamma, C) \equiv (\gamma', C')$ implies that $\gamma$ and $\gamma'$ are *region equivalent* in the sense of [5] and that $C \equiv C'$ in the sense of Definition 4.6. However Definition 8.7 doesn't just involve comparing fractional parts among the clock values in $C$, and separately among the clock values in $\gamma$: the fifth clause compares *between* values in $\gamma$ and values in $C$. This is essential for $\equiv$ to be a congruence with respect to the time-successor operation, as the following example shows.

**Example 8.8.** Let $(\gamma, C)$ and $(\gamma, C')$ be $\mathcal{B}$-$\mathcal{A}^c$ configurations, with $\gamma = (s, (1.1, 0.6))$, $C = \{(s_1, 0.5)\}$ and $C' = \{(s_1, 0.7)\}$. Note that $C \equiv C'$ (cf. Definition 4.6) and so, without the final clause in Definition 8.7, we would have $(\gamma, C) \equiv (\gamma, C')$. But it is clearly the case that $next(\gamma, C) \not\equiv next(\gamma, C')$. In fact, $next(\gamma, C)$ has the form $(\gamma', D)$ for some $\gamma'$ with $D = \{(s_1, 0.9)\}$, while $next(\gamma, D')$ has the form $(\eta, D')$ for some $\eta$ with $D' = \{(s_1, 1)\}$.

**Lemma 8.9** (Bisimulation Lemma). *Suppose that $(\gamma, C)$ and $(\eta, D)$ are $\mathcal{B}$-$\mathcal{A}^c$ configurations such that $(\gamma, C) \equiv (\eta, D)$. Then for each transition $(\gamma, C) \xrightarrow{\alpha} (\gamma', C')$, with $\alpha \in \Sigma \cup \{\varepsilon\}$, there exists a configuration $(\eta', D')$ such that $(\eta, D) \xrightarrow{\alpha} (\eta', D')$ and $(\gamma', C') \equiv (\eta', D')$.*

*Proof.* The proof is almost identical to that of Lemma 4.7. □

**Proposition 8.10.** *If configuration $(\gamma, C)$ is reachable from the initial configuration in $\mathcal{T}_{\mathcal{B}, \mathcal{A}^c}$, then there is a rational configuration $(\gamma', C')$, with $(\gamma, C) \equiv (\gamma', C')$, such that $(\gamma', C')$ is reachable from the initial configuration in $\mathcal{W}_{\mathcal{B}, \mathcal{A}^c}$.*

*Proof.* The proof is almost identical to that of Proposition 4.9. □

To complete the correspondence with Section 4, it remains to show that $\mathcal{W}_{\mathcal{B}, \mathcal{A}^c}$ is a wsts. As we now explain, this requires a slight variation of the construction used in Proposition 4.16.

Suppose that $\mathcal{A}$ has set of locations $S$ and that $\mathcal{B}$ has set of locations $T$, where $S$ and $T$ are disjoint. Define a finite alphabet $\Lambda$ to be the set of non-empty subsets of $((T \times \{1, \ldots, n\}) \cup S) \times REG$, where $REG$ is the set of clock regions as defined in Subsection 4.1. Following Definition 4.11, an *abstract $\mathcal{B}$-$\mathcal{A}^c$-configuration* is a finite word over $\Lambda$.

We reuse the abstraction function $H$ from Section 4 to map $\mathcal{B}$-$\mathcal{A}^c$-configurations to abstract configurations as follows: map a configuration $((s, \mathbf{v}), C)$ of $\mathcal{T}_{\mathcal{B}, \mathcal{A}^c}$ to the word $H(\{((s, 1), v_1), \ldots, ((s, n), v_n)\} \cup C) \in \Lambda^*$. From this word we can reconstruct all clock values in $((s, \mathbf{v}), C)$ up to the nearest integer and also the relative order of the fractional parts of the clocks. As in Proposition 4.13 this observation implies that the kernel of $H$ agrees with the notion of equivalence of $\mathcal{B}$-$\mathcal{A}^c$-configurations, that is, $H(\gamma, C) = H(\gamma', C')$ implies $(\gamma, C) \equiv (\gamma', C')$.

**Proposition 8.11.** *Define a quasi-order on $\mathcal{B}$-$\mathcal{A}^c$-configurations by $(\gamma, C) \preccurlyeq (\gamma', C')$ iff $H(\gamma, C) \sqsubseteq H(\gamma', C')$, where $\sqsubseteq$ refers to the subword order on $\Lambda^*$. Then $\mathcal{W}_{\mathcal{B}, \mathcal{A}^c}$ is a wsts when equipped with this quasi-order.*

*Proof.* The proof is almost identical to that of Proposition 4.16. □

**Theorem 8.12.** *Let $\mathcal{B}$ denote an Alur-Dill automaton, and $\mathcal{A}$ a one-clock alternating automaton in which every state is accepting. Then the language inclusion problem '$L_\omega(\mathcal{B}) \subseteq L_\omega(\mathcal{A})$?' is decidable.*

*Proof.* The inclusion $L_\omega(\mathcal{B}) \subseteq L_\omega(\mathcal{A})$ holds iff it is not possible to reach a doomed state from the initial state in $\mathcal{W}_{\mathcal{B},\mathcal{A}^c}$. Now the set of doomed states in $\mathcal{W}_{\mathcal{B},\mathcal{A}^c}$ is trivially downward-closed with respect to the monotone domination order (recall that $(\gamma, C)$ is doomed only if $C = \emptyset$). The set of doomed states is also decidable: to decide doom of $(\gamma, \emptyset)$ we have to check whether $\mathcal{B}$ can accept a non-Zeno timed word starting from $\gamma$. This last problem is essentially the language-emptiness problem for Alur-Dill automata over infinite timed words, which is well-known to be decidable—see [5]. Theorem 4.15 now yields a decision procedure for the language inclusion question '$L_\omega(\mathcal{B}) \subseteq L_\omega(\mathcal{A})$?'. □

**Corollary 8.13.** *The model-checking problem for Safety MTL over infinite words is decidable: given an Alur-Dill automaton $\mathcal{B}$ and a Safety MTL formula $\varphi$, there is an algorithm to decide whether or not $L_\omega(\mathcal{B}) \subseteq L_\omega(\varphi)$.*

*Proof.* Apply Theorem 8.12 in case $\mathcal{A} = \mathcal{A}_\varphi^{safe}$, using the result of Proposition 8.2 that $L_\omega(\varphi) = L_\omega(\mathcal{A}_\varphi^{safe})$. □

## 9. Conclusion

In this paper, we have shown that Metric Temporal Logic is decidable over finite timed words in its standard dense-time, point-based semantics, with non-primitive recursive complexity. Over infinite words, we have shown that the important safety fragment of Metric Temporal Logic can be model checked.

To prove the decidability results above, we introduced the class of timed alternating automata, and showed that the language-emptiness problem for one-clock timed alternating automata over finite words is decidable. In the words of [21], one-clock timed alternating automata constitute a *fully decidable* specification formalism for timed languages in that they are closed under all Boolean operations and language emptiness is decidable. In contrast to Alur-Dill timed automata, one-clock timed alternating automata do not admit finite untimed quotients. In fact, it is straightforward to define a one-clock timed alternating automaton $\mathcal{A}$ such that the untimed language obtained from $L_f(\mathcal{A})$ (by forgetting all timestamps) is the classic non-regular language $\{a^n b^m : 0 \leq n \leq m\}$. Reflecting this fact, the termination proof for our language emptiness algorithm used a well-quasi-order derived from Higman's Lemma.

The focus of this paper has exclusively been on MTL over finite words. Recently we have obtained both positive and negative decidability results for MTL over infinite words. In particular, we have shown that the satisfiability problem for Safety MTL is decidable [31], whereas the satisfiability problem for MTL is undecidable [30]. Thus restricting to safety properties is crucial to obtaining decidability.

**Acknowledgements.** We thank Tom Henzinger for clarifying some of the undecidability results for MTL, and for asking about the relationship between single-clock timed automata and real-time temporal logics. We also thank the anonymous referees for their comments, which significantly improved the presentation of the paper.